\documentclass[%
 reprint,
 superscriptaddress,
 aps,
 pra
]{revtex4-2}

\usepackage{graphicx}
\usepackage{dcolumn}
\usepackage{bm}
\usepackage{booktabs,makecell,tabularx}
\usepackage{multirow}
\usepackage{setspace}
\usepackage{amsfonts,amsmath,amssymb,amsthm}
\usepackage{hyperref}
\usepackage{natbib}
\usepackage{array}
\usepackage{rotating}
\usepackage{chngcntr}
\usepackage{rotating}
\usepackage{chngcntr}
\usepackage{xcolor}
\usepackage{tikz}
\usepackage{pgfplots}
\usepgfplotslibrary{groupplots}
\pgfplotsset{compat=1.18}

\counterwithout{figure}{section}

\newtheorem{theorem}{Theorem}
\hypersetup{
    pdfauthor = UnnamedOrange,
    colorlinks = true,
    linkcolor = blue,
    anchorcolor = red,
    citecolor = blue,
    urlcolor = blue}

\definecolor{journalblue}{RGB}{31,78,121}
\definecolor{journalbrick}{RGB}{176,74,58}

\begin{document}

\preprint{APS/123-QED}

\title{Pareto Front Engineering of Dynamical Sweet Spots in Superconducting Qubits}

\author{Zhen Yang}
\affiliation{Institute of Fundamental and Frontier Sciences, University of Electronic Science and Technology of China, Chengdu, 611731, China}
\affiliation{Key Laboratory of Quantum Physics and Photonic Quantum Information, Ministry of Education,
 University of Electronic Science and Technology of China, Chengdu, 611731, China}

\author{Shan Jin}
\affiliation{Institute of Fundamental and Frontier Sciences, University of Electronic Science and Technology of China, Chengdu, 611731, China}
\affiliation{Key Laboratory of Quantum Physics and Photonic Quantum Information, Ministry of Education,
 University of Electronic Science and Technology of China, Chengdu, 611731, China}
\affiliation{Yangtze Delta Industrial Innovation Center of Quantum Science and Technology, Suzhou 215000, China}

\author{Yajie Hao}
\affiliation{Institute of Fundamental and Frontier Sciences, University of Electronic Science and Technology of China, Chengdu, 611731, China}
\affiliation{Key Laboratory of Quantum Physics and Photonic Quantum Information, Ministry of Education,
 University of Electronic Science and Technology of China, Chengdu, 611731, China}

\author{Guangwei Deng}
\email{gwdeng@uestc.edu.cn}
\affiliation{Institute of Fundamental and Frontier Sciences, University of Electronic Science and Technology of China, Chengdu, 611731, China}
\affiliation{Key Laboratory of Quantum Physics and Photonic Quantum Information, Ministry of Education,
 University of Electronic Science and Technology of China, Chengdu, 611731, China}

\author{Xiu-Hao Deng}
\email{dengxiuhao@iqasz.cn}
\affiliation{International Quantum Academy, Shenzhen, 518048, China}
\affiliation{Shenzhen Branch, Hefei National Laboratory, Shenzhen, 518048, China}

\author{Re-Bing Wu}
\email{rbwu@tsinghua.edu.cn}
\affiliation{%
    Center for Intelligent and Networked Systems, Department of Automation, Tsinghua University, Beijing 100084, China}

\author{Xiaoting Wang}
\email{xiaoting@uestc.edu.cn}
\affiliation{Institute of Fundamental and Frontier Sciences, University of Electronic Science and Technology of China, Chengdu, 611731, China}
\affiliation{Key Laboratory of Quantum Physics and Photonic Quantum Information, Ministry of Education,
 University of Electronic Science and Technology of China, Chengdu, 611731, China}

\date{\today}

\begin{abstract}
Operating superconducting qubits at dynamical sweet spots (DSSs) suppresses decoherence from low-frequency flux noise. A key open question is how long coherence can be extended under this strategy and what fundamental limits constrain it. Here we introduce a fully parameterized, multi-objective periodic-flux modulation framework that simultaneously optimizes energy relaxation ($T_1$) and pure dephasing ($T_\phi$), thereby quantifying the tradeoff between them. For fluxonium qubits with realistic noise spectra, our method enhances $T_\phi$ by a factor of 3-5 compared with existing DSS strategies while maintaining $T_1$ in the hundred-microsecond range. We further prove that, although DSSs eliminate first-order sensitivity to low-frequency noise, relaxation rate cannot be reduced arbitrarily close to zero, establishing an upper bound on achievable $T_1$. As applications, we design {leakage-suppressed} single- and two-qubit control protocols at the optimized operating points and numerically demonstrate high-fidelity gates. These results establish a general and useful framework for Pareto-front engineering of DSSs that substantially improves coherence and gate performance in superconducting qubits.
\end{abstract}

\maketitle

\section{Introduction}

Superconducting qubits are among the most advanced platforms for quantum computation, with the transmon qubit being widely studied in both theory and experiment~\cite{Nakamura1999_CPB, Mooij1999_SC, wallraff2004_SC, Blais2004_cQED, schuster2007_cQED, Koch2007_Transmon, houck2012_chip, riste2013_feedback, WangHaoHua2019_PRL, arute2019_SC, Blais2021_rmp, ZhuXiaobo2021_PRL, LuYan2023_breakeven, kim2023_127, bravyi2024_HighThreshold, google2025quantum, putterman2025_hardware}. In parallel, the fluxonium qubit has emerged as a promising alternative, offering long coherence times and high anharmonicity~\cite{Vladimir2009_Fluxonium, Nguyen2019_Fluxonium, Nguyen2022_Processor}. Nevertheless, preserving quantum coherence remains a central challenge, because qubits inevitably couple to their environments. Extending coherence times and mitigating noise-induced errors are therefore critical goals for both fault-tolerant quantum computing and near-term noisy intermediate-scale quantum processors~\cite{Terhal2015_QEC, Fowler2012_SurfaceCode, Preskill2018_NISQ, Temme2017_ErrorMitigation}. 
In fluxonium devices, dominant decoherence mechanisms include flux noise~\cite{Yoshihara2006_Decoherence} and dielectric loss~\cite{Martinis2005_Dielectric}. Several strategies have been developed to improve coherence, including operating at static sweet spots~\cite{Ithier2005_Decoherence,Koch2007_CooperPair,Nguyen2019_Coherence,Zhang2021_FluxControl,Sete2017_Tunable}, applying dynamical decoupling techniques~\cite{Viola1999_Decoupling,Khodjasteh2005_Decoupling,Cywinski2008_Dephasing,Uhrig2007_Optimization,Pokharel2018_Fidelity}, and advancing materials and fabrication techniques~\cite{Martinis2005_Dielectric, Wang2015_Dielectric}. More recently, periodic flux modulation has been proposed to combine the advantages of static sweet spots and dynamical decoupling. This approach gives rise to dynamical sweet spots (DSSs), where the first-order sensitivity to low-frequency flux noise is suppressed, thereby reducing pure dephasing and improving coherence in periodically driven systems. DSSs have since been explored extensively in both theory and experiment, demonstrating substantial robustness to noise~\cite{didier2019_DSS2, Valery2022_DSS, Huang2021_SweetSpots, Mundada2020_Floquet, Gandon2022_Floquet, Cheng2022_Coherence, Joachim2025_TwoTone, frees2019_DSS, guo2018_DSS, pirkkalainen2013_DSS}. In transmons, they were introduced via single-tone parametric flux modulation~\cite{didier2019_DSS2}, with two-tone drives later improving gate performance in experiments~\cite{Valery2022_DSS}. In fluxonium, Huang et al.~\cite{Huang2021_SweetSpots} established Floquet-engineered continuous DSS families that support control and readout, with experiments reporting up to a 40-fold increase in coherence at such DSSs~\cite{Mundada2020_Floquet}. Subsequent studies have further expanded the DSS toolbox: longitudinal readout eliminates the adiabatic ramp in dispersive schemes~\cite{Gandon2022_Floquet}; tunable-complex-amplitude modulation~\cite{Cheng2022_Coherence} and commensurate two-tone modulation~\cite{Joachim2025_TwoTone} further extend the control space and improve coherence times; two-tone modulation provides access to double-DSS operating points, with analytic formulas for their locations that are useful for parameter selection~\cite{Dominic2025_SweetSour}. 
Overall, DSSs have evolved from single-tone transmon implementations to Floquet-engineered fluxonium families, expanding control capabilities and boosting robustness. 

Despite this progress, important limitations remain. At static sweet spots, qubits typically exhibit longer $T_\phi$ but shorter $T_1$ compared to points far from the sweet spot, reflecting an inherent trade-off between these two figures of merit. A similar competition arises in DSSs: the system parameters that maximize $T_\phi$ generally do not coincide with those that maximize $T_1$~\cite{Cheng2022_Coherence}. Moreover, existing DSS schemes often restrict the modulation to predefined forms, such as single-tone or two-tone drives~\cite{Huang2021_SweetSpots, Mundada2020_Floquet, Cheng2022_Coherence, Gandon2022_Floquet, Joachim2025_TwoTone}. These constraints naturally raise the question: can coherence times $T_1$ and $T_\phi$ be optimized simultaneously under general periodic flux control, and what fundamental limits govern their maximum values? 

To overcome the limitations of existing DSS approaches and to systematically determine the optimal coherence times, we develop a fully parameterized multi-objective periodic-flux modulation framework for fluxonium qubits. This framework generalizes beyond conventional single- and two-tone drives by allowing arbitrary periodic flux modulations, {parameterized through their Fourier coefficients}, thereby enlarging the search space for waveform design. Within this setting, we perform a multi-objective optimization of $T_1$ and $T_\phi$ and map out the resulting $T_1$-$T_\phi$ Pareto front (PF), which makes the trade-off explicit and identifies optimal operating points. For realistic experimental parameters, the optimal solutions on this PF extend $T_\phi$ by factors of 3-5 compared with existing DSS strategies, while maintaining $T_1$ in the hundred-microsecond regime. By analyzing the PF, we find that relaxation times $T_1$ cannot be increased without bound {within the DSS regime under the specified $\sigma_z$-coupled noise model}. 

{As applications, we select a representative DSS operating point on the PF and demonstrate gate operations by superimposing additional charge-drive control pulses on the optimized periodic flux modulations. The gates are optimized and validated in a multilevel Floquet rotating frame, so that leakage outside the Floquet computational subspace is explicitly included in the fidelity evaluation. In a validation model with truncation \(N=12\), we obtain a single-qubit \(X\) gate with average fidelity \(F_{\rm avg}=99.99\%\) and average leakage \(L_{\rm avg}=9.35\times10^{-6}\) for a gate duration of \(11~{\rm ns}\). For the two-qubit \(\sqrt{i\text{SWAP}}\) gate, we obtain \(F_{\rm avg}=99.92\%\) and \(L_{\rm avg}=7.18\times 10^{-4}\) for a gate duration of \(34~{\rm ns}\). These results show that the optimized DSS operating points can support high-fidelity gates even when multilevel leakage is explicitly included in the validation. We further evaluate gate fidelities in an open-system simulation, obtaining \(99.99\%\) for the single-qubit \(X\) gate and \(99.90\%\) for the two-qubit \(\sqrt{i\text{SWAP}}\) gate.}

Unless otherwise stated, we focus on fluxonium qubits throughout this work, though similar techniques are also applicable to transmon devices. This paper is organized as follows. Section~\ref{sec:hamiltonian} introduces the dynamics of fluxonium qubits under general periodic modulation and establishes the connection between tunable parameters and decoherence rates. Section~\ref{sec:optimization} describes the multi-objective numerical optimization procedure to simultaneously optimize $T_1$ and $T_\phi$, and reports the $T_1$-$T_\phi$ Pareto front, {including the identification and construction of DSS as well as the model-dependent upper bound on $T_1$}. Section~\ref{sec:gates} demonstrates the realization of high-fidelity quantum gates at optimal DSS operating points.

\section{Fluxonium qubits under general Periodic Modulation}\label{sec:hamiltonian}

Under general periodic modulation, the dynamics of fluxonium qubits are well described within the Floquet framework.  
This formalism allows us to analyze the system-bath interaction and to derive the associated decoherence rates in the presence of low-frequency $1/f$ flux noise and dielectric loss.

\subsection{System Hamiltonian}

The Hamiltonian of the fluxonium system is given by~\cite{Manucharyan2009_Fluxonium}
\begin{equation}
    H_q(t)=4 E_C \hat{n}^2+\frac{1}{2} E_L\left[\hat{\varphi}+\phi_{\mathrm{ext}}(t)\right]^2-E_J \cos \hat{\varphi},
\end{equation}
where $E_C$, $E_L$, and $E_J$ denote the capacitive, inductive, and Josephson energies, respectively. $\hat{n} = Q / (-2e)$ represents the Cooper-pair number operator, while $\hat{\varphi} = 2\pi \Phi / \Phi_0$ is the dimensionless flux operator. Here, $\Phi_0 = h / (2e)$ is the magnetic flux quantum, where $h$ is Planck's constant. The charge operator $Q$ and the flux operator $\Phi$ satisfy the canonical commutation relation $[\Phi, Q] = i\hbar$, which leads to $[\hat{n}, \hat{\varphi}] = i$. In this work, we discuss the most general form of the external periodic flux drive:
\begin{equation}
    \phi_{\text{ext}}(t) = \phi_{\mathrm{dc}} + \phi_{\mathrm{ac}}P(t),
\end{equation}
{where \(\phi_{\mathrm{dc}}\) is the static DC component, \(\phi_{\mathrm{ac}}\) sets the overall modulation scale, and \(P(t)\) is the periodic AC component with frequency \(\omega_d\). The real-valued periodic function \(P(t)\) can be expanded in Fourier series as \(P(t)=\sum_n p_n e^{in\omega_d t}\). The Fourier coefficients \(p_n\in\mathbb{C}\) are treated as trainable parameters, subject to the Hermiticity condition \(p_{-n}=p_n^\ast\) to ensure that \(P(t)\) is real. Consequently, the Hamiltonian is time-periodic, satisfying \(H_q(t)=H_q(t+T)\) with period \(T=2\pi/\omega_d\). Because the zeroth Fourier component \(p_0\) contributes to the DC part of the waveform, it is convenient to distinguish the parametrization variables in Eq.~(2) from the corresponding physical flux quantities. 
We therefore rewrite the external flux as $\phi_{\mathrm{ext}}(t)=\tilde{\phi}_{\mathrm{dc}}+\tilde{\phi}_{\mathrm{ac}}\tilde P(t)$,
where $\tilde{\phi}_{\mathrm{dc}}=\phi_{\mathrm{dc}}+\phi_{\mathrm{ac}}p_0$ is the physical DC flux bias; $\tilde{\phi}_{\mathrm{ac}}=\phi_{\mathrm{ac}}\max_t|P(t)-p_0|\equiv\phi_{\mathrm{ac}}\eta_P$ is the physical peak AC modulation amplitude; and $\tilde P(t)=\left(P(t)-p_0\right)/\eta_P$ is the normalized waveform satisfying $\max_t|\tilde P(t)|=1$.

At a physical DC flux bias of \({\phi}_{\mathrm{ext}}=\pi\), the fluxonium system operates at the so-called static sweet spot, where the qubit energy splitting is first-order insensitive to low-frequency flux noise~\cite{Huang2021_SweetSpots}. In contrast, away from this bias point (i.e., at an off-sweet-spot such as \(\tilde{\phi}_{\mathrm{dc}}=1.03\pi\)), the system regains flux sensitivity, which can be exploited for controlled modulation at the expense of increased susceptibility to noise.}

{
Restricting the dynamics to the lowest two eigenstates at the half-flux sweet spot, $\{|\tilde g\rangle,|\tilde e\rangle\}$, we define the two-level projector $\Pi_2=|\tilde g\rangle\langle\tilde g|+|\tilde e\rangle\langle\tilde e|$. In this static sweet-spot energy basis, the unrotated Pauli operators are defined as $\tilde{\sigma}_z=|\tilde e\rangle\langle\tilde e|-|\tilde g\rangle\langle\tilde g|,\;\tilde{\sigma}_x=|\tilde g\rangle\langle\tilde e|+|\tilde e\rangle\langle\tilde g|$. With a suitable phase choice of $|\tilde g\rangle$ and $|\tilde e\rangle$, the nontrivial part of the projected phase operator $\Pi_2\hat{\phi}\Pi_2$ is aligned with $\tilde{\sigma}_x$; the detailed projection is given in Supplementary Sec.~I.A. Following the rotated Pauli convention used in Refs.~\cite{Huang2021_SweetSpots,Cheng2022_Coherence}, we apply the fixed rotation $U_y=\exp\left(-i\frac{\pi}{4}\tilde{\sigma}_y\right)$ and relabel the rotated operators as $\tilde{\sigma}_z\rightarrow\sigma_x,\; \tilde{\sigma}_x\rightarrow-\sigma_z$. After dropping identity terms, the effective two-level Hamiltonian is therefore written as
\begin{equation}\label{Hq_2level}
H_q(t)=\frac{\Delta}{2}\sigma_x+\frac{B}{2}\sigma_z+AP(t)\sigma_z ,
\end{equation}
where $\Delta=\tilde E_e-\tilde E_g$ is the sweet-spot energy splitting, $B=2E_L(\varphi_{dc}-\pi)\tilde{\phi}_{ge}$, and $A=E_L\varphi_{ac}\tilde{\phi}_{ge},\; \tilde{\phi}_{ge}=|\langle\tilde g|\hat{\phi}|\tilde e\rangle|$. 
The same form is retained when using the normalized waveform parametrization, $\varphi_{\mathrm{ext}}(t)=\tilde{\phi}_{dc}+\tilde{\phi}_{ac}\tilde P(t)$.
Thus, in the Pauli convention used throughout this work, $\sigma_x$ denotes the static sweet-spot splitting axis, whereas $\sigma_z$ denotes the two-level representation of the projected phase operator, i.e., the system operator associated with flux coupling.
}

Since the effective Hamiltonian $H_q(t)$ contains an explicit time-periodic drive, the system’s dynamics are governed by the time-dependent Schrödinger equation with a periodic Hamiltonian. In this setting, Floquet theory~\cite{Kohler1997_Floquet} provides a natural framework for describing the solutions. Specifically, the wavefunctions can be expressed in the Floquet form
\begin{equation}
\left| \psi_{\pm}(t) \right\rangle = e^{-i \epsilon_{\pm} t} \left| \omega_{\pm}(t) \right\rangle ,
\end{equation}
where $\epsilon_{\pm}$ are the quasienergies, and $\left|{\omega}_{\pm}(t)\right\rangle$ are the corresponding Floquet states, which are periodic in time with the same period as $H_q(t)$. The quasienergies and Floquet states are obtained by solving the Floquet equation~\cite{Creffield2003_Floquet}
\begin{equation}
    \left[ H_q(t) - i \frac{\partial}{\partial t} \right] \left| \omega_\pm(t) \right\rangle = \epsilon_\pm \left| \omega_\pm(t) \right\rangle.
\end{equation}

By expanding the Hamiltonian and Floquet states in Fourier components, we have $H_q(t)=\sum_{n \in \mathbb{Z}} H_q^{[n]} e^{-i n \omega_d t}$ and $\left| \omega_\pm(t) \right\rangle=\sum_{k \in \mathbb{Z}}\left|\omega_\pm^{[k]}\right\rangle e^{-i k \omega_d t}$, and the operator $\left[ H_q(t) - i \frac{\partial}{\partial t} \right]$ can be recast as an infinite-dimensional Floquet matrix $\mathcal{H}_F$ in the basis $\left\{\left| \omega_\pm^{[k]}\right\rangle, k\in\mathbb{Z}  \right\}$~\cite{Cheng2022_Coherence}:
\begin{equation}\label{floquet_matrix}
    \mathcal{H}_F = \begin{pmatrix}
        \ddots                                 & \vdots                    & \vdots                & \vdots                    & \begin{sideways}$\ddots$\end{sideways} \\
        \cdots                                 & H_q^{[0,-1]} & H_q^{[-1]}            & H_q^{[-2]}                & \cdots                                 \\
        \cdots                                 & H_q^{[1]}                 & H_q^{[0,0]} & H_q^{[-1]}                & \cdots                                 \\
        \cdots                                 & H_q^{[2]}                 & H_q^{[1]}             & H_q^{[0,1]} & \cdots                                 \\
        \begin{sideways}$\ddots$\end{sideways} & \vdots                    & \vdots                & \vdots                    & \ddots
    \end{pmatrix}
\end{equation}
The off-diagonal blocks are \(H_q^{[n\neq 0]}=A p_n \sigma_z\), and the diagonal blocks read \(H_q^{[0,k]}=H_q^{[0]}-k\,\omega_d\,\mathbb I\) where $H_q^{[0]} = \frac{\Delta}{2} \sigma_x + \left( \frac{B}{2} + p_0 A \right) \sigma_z$. Diagonalizing $\mathcal{H}_F$ yields the quasienergies, corresponding to the pair of eigenvalues with the smallest absolute value, while the Floquet states are obtained from the associated eigenvectors.

\subsection{Noise and Decoherence Rates}

{
For completeness, we begin by reviewing the derivation of decoherence rates in a fluxonium qubit subject to environmental noise. The total Hamiltonian is
\begin{equation}
H(t)=H_q(t)+H_B+H_{\mathrm{int}} .
\end{equation}
The system-bath interaction is written as $H_{\mathrm{int}}=\sigma_z\beta$, where $\sigma_z$ is the same two-level representation of the projected phase operator used in Eq.~\eqref{Hq_2level}. For flux noise, this form follows from expanding the inductive energy to first order in the external-flux fluctuation; for dielectric loss, we follow the effective noise model of Refs.~\cite{Huang2021_SweetSpots,Cheng2022_Coherence}. Details of the projection and noise-coupling convention are given in Supplementary Sec.~I.B.}

To analyze the system dynamics, we move into the rotating frame defined by $U_q(t)\otimes U_B(t)$, where $U_q(t)=\mathcal{T}\exp[-i\int_0^tH_q(t')dt']$ and $U_B(t)=\exp(-iH_Bt)$. Here $\mathcal{T}$ is the time-ordering operator. In this frame, the interaction Hamiltonian becomes $H_{\mathrm{int}}(t)=\sigma_z(t)\beta(t)$, with $\sigma_z(t)=U_q^\dagger(t)\sigma_z U_q(t)$ and $\beta(t)=U_B^\dagger(t)\beta U_B(t)$, which encodes the system's time-dependent sensitivity to noise.

In the Floquet state basis $\{|\omega_\pm(t)\rangle\}$, we introduce the operator basis $\{\tau_j(t)\,|\,j=z,\pm\}$, where $\tau_z(t)=\left|\omega_{+}(t)\right\rangle\left\langle\omega_{+}(t)\right|-\left|\omega_{-}(t)\right\rangle\left\langle\omega_{-}(t)\right|, $ and $\tau_{+}(t)=\tau_{-}^{\dagger}(t)=\left|\omega_{+}(t)\right\rangle\left\langle\omega_{-}(t)\right|$.
In this representation, $\sigma_z(t)$ admits a Bohr-frequency decomposition, $\sigma_z(t) = \sum_{\omega=0, \pm\Omega} \sigma_z(\omega)$, 
with components  
$\sigma_z(0)=\tfrac{1}{2}\tau_z(0)\operatorname{Tr}[\sigma_z \tau_z(t)]$ and  
$\sigma_z(\pm \Omega)=\tau_\pm(0)e^{\pm i \Omega t}\operatorname{Tr}[\sigma_z\tau_\mp(t)]$.  
Here, $\Omega = \epsilon_+ - \epsilon_-$ denotes the quasienergy gap. 
The bath operator $\beta(t)$ determines the noise correlation function, $C(t)=\operatorname{Tr}[\beta(t)\beta(0)\rho_B]$, whose Fourier transform yields the noise spectral density $S(\omega)$. 

The reduced dynamics of the qubit are governed by the second-order time-convolutionless master equation~\cite{Suarez2025_nonmarkovn}:  
\begin{equation}
    \frac{d}{dt}\rho(t)
    = i\left[\rho(t),H_{LS}(t)\right]+ \sum_{\omega}\gamma(\omega,t)\,\mathcal{D}[\sigma_z(\omega)]\rho(t),
\end{equation}
where Lamb-shift Hamiltonian $H_{LS}(t)$ is $\sum_{\omega}\mathcal{S}(\omega,t)\,\sigma_z^\dagger(\omega)\sigma_z(\omega)$, dissipator $\mathcal{D}[c]\rho \equiv c\rho c^\dagger - \tfrac{1}{2}\{c^\dagger c,\rho\}$, and coefficients $S(\omega, t)=\operatorname{Im}[\Gamma(\omega, t)], \gamma(\omega, t)=2 \operatorname{Re}[\Gamma(\omega, t)]$ and $ \Gamma(\omega, t)=\int_0^t d s e^{i \omega s} C(s)$. Expanding the periodic coefficients in Fourier harmonics, $\operatorname{Tr}[\sigma_z\tau_j^\dagger(t)] = \sum_{k\in\mathbb{Z}} g_j^{[k]} e^{-i k \omega_d t}\,(j=z,\pm)$, the time-dependent decoherence rates take the form
\begin{equation}
\gamma_j(t) = \frac{1}{a_j} \int_{-\infty}^{+\infty} d\omega \left[ \sum_k \frac{t}{\pi}\,\mathrm{sinc}\!\left((\omega+\omega_{k,j})t\right) |g_j^{[k]}|^2 \right] S(\omega),
\end{equation}
where $a_z=4$, $a_\pm=1$, $\omega_{k,z}=-k\omega_d$, $\omega_{k,\pm}=\pm\Omega-k\omega_d$, and Fourier coefficients
\begin{equation}\label{floquet_gamma}
g_j^{[k]} = \frac{1}{b_j T}\int_0^T dt\, \operatorname{Tr}[\sigma_z\tau_j^\dagger(t)] e^{i k \omega_d t},
\end{equation}
with $b_z=2,b_\pm=1$. For sufficiently long time ($t\gg 1/\omega_{k,j}$), the rates converge to time-independent values,
\begin{equation}
 \gamma_z=\sum_k\left|g_z^{[k]}\right|^2 S\left(k \omega_d\right),\; \gamma_{ \pm}=\sum_k\left|g_{ \pm}^{[k]}\right|^2 S\left(k \omega_d \mp {\Omega}\right).
\end{equation}
In practice, decoherence in fluxonium qubits is often dominated by $1/f$ flux noise and dielectric loss noise~\cite{Manucharyan2012_Phaseslips,Klimov2018_Energy,Nguyen2019_Coherence}, with spectral density
\begin{equation}\label{S}
S(\omega) = A_f^2 \Big|\tfrac{2\pi}{\omega}\Big| + \kappa(\omega,\mathcal{T})A_d\Big(\tfrac{\hbar\omega}{2\pi}\Big)^2,
\end{equation}
where $A_f$ and $A_d$ are the respective amplitudes. The thermal factor is $\kappa(\omega,\mathcal{T})=\tfrac{1}{2}\left|\coth\!\left(\tfrac{\hbar\omega}{2k_B\mathcal{T}}\right)+1\right|$. To regularize the infrared divergence of $1/f$ noise, we introduce cutoffs $\omega_{\mathrm{ir}}=1\,\mathrm{Hz}$ and $\omega_{\mathrm{uv}}=3\,\mathrm{GHz}$ following Ref.~\cite{Groszkowski2018_noise}. Under this model, the pure dephasing rate $\gamma_z$ and energy relaxation rate $\gamma_\pm$ are~\cite{Cheng2022_Coherence}:
\begin{equation}\label{gamma}
\begin{split}
\gamma_z &=  |g_z^{[0]}| A_f \sqrt{2|\ln(\omega_{\mathrm{ir}} t_m)|} 
+  \sum_{k\neq 0} |g_z^{[k]}|^2 S(k\omega_d), \\
\gamma_\pm &= \sum_{k\in\mathbb{Z}}  |g_\pm^{[k]}|^2 S(k\omega_d\mp\Omega) 
,
\end{split}
\end{equation}
where $\sqrt{|\ln(\omega_{\mathrm{ir}}t_m)|}\!\approx\!4$ since $\omega_{\mathrm{ir}}$ is much smaller than the inverse timescale $t_m$~\cite{Kou2017_Molecule,Sete2017_Tunable}. 
{The derivation above is written in the two-level Floquet basis used for the main optimization. 
In Supplementary Sec.~III.A we derive the corresponding \(N\)-level full-fluxonium version, in which the same rates are evaluated from the Floquet modes of the truncated full Hamiltonian. This multilevel formulation is used to verify that higher fluxonium levels introduce quantitative corrections to the decoherence rates.}

Finally, based on $\gamma_\pm$ and $\gamma_z$, one can define the two types of coherence times, the energy relaxation time $T_1$ and pure dephasing time $T_\phi$:
\begin{align}
T_1 = \frac{1}{\gamma_+ + \gamma_-}\equiv\frac{1}{\gamma_1}, \, T_\phi = \frac{1}{\gamma_z}
\end{align}
where $\gamma_1 = \gamma_+ + \gamma_-$. 
Operating the fluxonium system at DSSs can generally improve coherence time. The next question is how to locate the DSSs given the general form of the periodic flux drive $\phi_{\text{ext}}(t) = \phi_{\mathrm{dc}} + \phi_{\mathrm{ac}}P(t)$.

\section{Improving Coherence Times via Pareto-Front Optimization}\label{sec:optimization}

For quantum computing applications, we aim to maximize both coherence times $T_1$ and $T_\phi$, which is equivalent to minimizing both energy relaxation rate $\gamma_1$ and dephasing rate $\gamma_z$. In this section, under the most general periodic flux drive  
\begin{equation}
    \phi_{\text{ext}}(t) = \phi_{\mathrm{dc}} + \phi_{\mathrm{ac}} \sum_n p_n e^{in\omega_d t},
\end{equation}
we show how to simultaneously optimize both $\gamma_1$ and $\gamma_z$ under the DSS framework.

\subsection{Joint optimization of $\gamma_1$ and $\gamma_z$}

Under the general periodic flux modulation function $\phi_{\text{ext}}(t)$, the Floquet states $|\omega_j(t)\rangle$ depend explicitly on the modulation parameters $\{p_n\}$, and different parameter choices lead to distinct decoherence behaviors. We aim to solve the following bi-objective optimization problem:
\begin{equation}\label{target}
\min_{\{p_n\}}\; \big(\gamma_1,\;\gamma_z\big),
\end{equation}
subject to the normalization condition
\begin{equation}\label{trade_off_ac}
    \sum_{k \in \mathbb{Z}}\left( \tfrac{1}{2}\left|g_{+}^{[k]}\right|^2 + \tfrac{1}{2}\left|g_{-}^{[k]}\right|^2 + \left|g_z^{[k]}\right|^2 \right) = 1.
\end{equation}

In a general multi-objective optimization problem, there is no single solution that simultaneously minimizes all objectives. Improving one objective often worsens another. The Pareto set is defined as the collection of parameter choices where neither $\gamma_1$ nor $\gamma_z$ can be improved without sacrificing the other. 
Accordingly, the Pareto front (PF) is the set of non-dominated objective values in the \((\gamma_1,\gamma_z)\) plane, representing the optimal trade-offs between energy relaxation and pure dephasing.
Consequently, solving a multi-objective optimization problem is often formulated as identifying and calculating the PF.

We note that gradient-based optimization methods may be impractical here: although they often converge faster, they require gradients with respect to $\{p_n\}$. Analytic gradients are difficult to obtain because each evaluation of the objective needs diagonalization of the truncated Floquet matrix, so numerical gradient estimates (e.g., finite differences) become computationally expensive as the parameter dimension grows.
Moreover, the normalization condition Eq.~(\ref{trade_off_ac}) implies an inherent trade-off between $\gamma_1$ and $\gamma_z$. These considerations motivate the use of heuristic, population-based multi-objective algorithms, such as genetic algorithms~\cite{Deb2002_NSGA2}.

Our approach begins by randomly initializing a population of $M$ candidate solutions, each associated with an objective vector $(\gamma_1,\gamma_z)$. At each generation, this population serves as the parent pool to produce $M$ 
\begin{minipage}{\linewidth}
    \vspace{2pt}
    \hrule height 0.7pt \vspace{1pt}\hrule height 0.7pt
    \vspace{4pt}
    \noindent
    \textbf{Algorithm 1.} \textit{Multi-objective optimization of decoherence rates} \par
    \vspace{2pt}\hrule height 0.5pt \vspace{4pt}

    \textbf{Stage I — Evolutionary search} \par
    \hspace{1em}\textbf{S1:} Initialize a population of \(M\) individuals with random parameters \(\{p_n\}\). \par
    \hspace{1em}\textbf{S2:} For each generation do: \par
    \hspace{2em}{S2.1:} Evaluate all individuals to obtain objective vectors $(\gamma_1,\gamma_z)$. \par
    \hspace{2em} {S2.2:} Use the entire current population as the parent pool. \par
    \hspace{2em} {S2.3:} Apply crossover and mutation to generate $M$ offspring. \par
    \hspace{2em} {S2.4:} Merge parents and offspring (size \(2M\)); apply an environmental-selection procedure to retain \(M\) individuals for the next generation. \par
    \hspace{2em} {S2.5:} Check stopping condition. \par
    \hspace{1em}\textbf{S3:} Output the non-dominated solutions of the final population as the run-level PF. \par
    \vspace{4pt}
    \textbf{Stage II — PF aggregation} \par
    \hspace{1em}\textbf{S4:} Repeat Stage I to obtain multiple run-level PFs. \par
    \hspace{1em}\textbf{S5:} Aggregate all run-level PF into a combined set. \par
    \hspace{1em}\textbf{S6:} Perform non-dominated sorting and remove any dominated solution. \par
    \hspace{1em}\textbf{S7:} Output the non-dominated solution as the aggregated PF. \par
    \vspace{4pt}
    \hrule height 0.7pt \vspace{1pt}\hrule height 0.7pt
\end{minipage}
\\[12pt]
offspring through variation, crossover, and mutation. After variation, an environmental selection step chooses which individuals survive to the next generation.
The choice of selection mechanism determines the class of multi-objective optimization algorithm used. This generate-combine-select loop is iterated until a stopping condition is met (e.g., a fixed number $N$ of generations, or convergence of a performance metric). At termination, the algorithm returns the set of non-dominated solutions from the final population as its PF, where dominance is defined as: solution $A$ dominates $B$ if $A$ is no worse in every objective and strictly better in at least one. 

In general, an analytical expression for the PF is generally unavailable. To obtain a more robust approximation, we run different multi-objective algorithms and merge their outputs to produce a collection of candidate PFs. Then we aggregate the non-dominated solutions from these fronts, and remove any solution that is dominated by another solution. 
The remaining non-dominated solutions aggregate into a curve, called the aggregated PF, which we treat as the final numerical solution set of the multi-objective optimization problem.
The full procedure is summarized in Algorithm~1. 
Specifically, we draw on a pool of over 30 multi-objective optimization algorithms for Algorithm~1, including NSGA2~\cite{Deb2002_NSGA2}, SPEA2~\cite{Zitzler2001_SPEA2}, tDEA~\cite{Li2018_TwoArchive}, IBEA~\cite{Zitzler2004_IBEA}, HypE~\cite{Bader2011_HypE}, MOEA/D\cite{Zhang2007_MOEAD}, ENS-MOEA/D~\cite{Zhao2012_Ensemble}, etc.
For implementation details of the algorithms and the corresponding simulation we refer the reader to Ref.~\cite{EMOC_colalab}. In our simulations, each algorithm is considered to converge when both the number and position of the non-dominated solutions on the PF show no significant change between successive generations. Here, we adopt numerical stopping criteria rather than relying on analytic convergence, due to the absence of analytical expressions for the Pareto front. To ensure a fair comparison across algorithms, the number of iterations is fixed at $N=2000$ for all runs.

\begin{figure}[tb]
    \centering
    \includegraphics[width=1.0\linewidth]{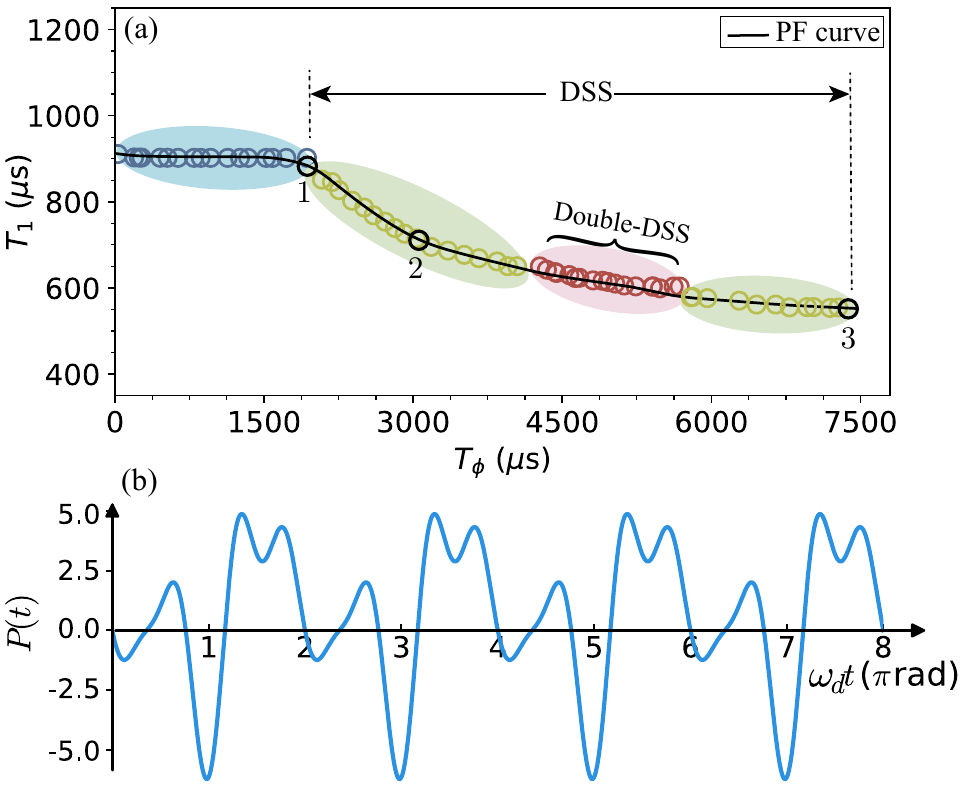}
    \caption{
    (a) Aggregated PF with Fourier truncation \(n=4\), obtained by aggregating the run-level PFs from ENS-MOEA/D, HypE, SPEA2, and tDEA followed by a final non-dominated sort. 
    Working Point 1 marks the onset of the DSS region, where \(|g_z^{[0]}|<10^{-4}\). 
    The red segment labeled ``Double-DSS'' marks the PF seed region used for constructive double-DSS refinement. 
    The final double-DSS points are obtained from these PF seeds by fixing the normalized waveform shape and drive frequency and solving \(F_{dc}=0\) and \(F_{ac}=0\) in the local \((\tilde{\varphi}_{dc},\tilde{\varphi}_{ac})\) plane.
    (b) Time-domain modulation for Working Point 2: \(P(t)=\sum_n p_n e^{in\omega_d t}\), defining \(\phi_{\mathrm{ext}}(t)=\phi_{\mathrm{dc}}+\phi_{\mathrm{ac}}P(t)\).
    }
    \label{Result_PF}
\end{figure}

Different optimization algorithms explore complementary regions of the objective space. Among them, ENS-MOEA/D, HypE, SPEA2 and tDEA collectively produce a more comprehensive PF. We therefore combine the run-level PFs from these four algorithms to obtain the aggregated PF via a final non-dominated sorting. The resulting PF, shown in Fig.~\ref{Result_PF}(a), illustrates the trade-off between the two optimization objectives $\gamma_1$ and $\gamma_z$, corresponding to the trade-off between $T_1$ and $T_\phi$. Based on its relation to the DSSs, 
{
the aggregated PF can be divided into two distinct segments: a DSS region and a non-DSS region. 
As discussed in the next subsection, the DSS region is located predominantly to the right of Working Point~1 in Fig.~\ref{Result_PF}(a). 
Moreover, within the DSS region, a subset of PF points can be further refined to construct double-DSS operating points. }
More detailed results for the aggregated PF are provided in Supplementary Sec.~II.A.

The aggregated PF indicates that our arbitrary-periodic flux-modulation approach attains longer coherence times than previously reported DSS schemes. We mark three representative working points on the PF in Fig.~\ref{Result_PF}(a). {
Their corresponding Fourier parameters are provided in Supplementary Table~S1, together with the physical quantities \(\tilde{\phi}_{\mathrm{dc}}\), \(\tilde{\phi}_{\mathrm{ac}}\), and \(\omega_d\).} At Working Point 2 the energy relaxation time $T_1$ is comparable to the $718\ \mu\mathrm{s}$ reported for the tunable-complex-amplitude scheme~\cite{Cheng2022_Coherence}, while the dephasing time reaches $T_\phi=3035\ \mu\mathrm{s}$. At Working Point 3 the dephasing time attains $T_\phi=7398\ \mu\mathrm{s}$, roughly three- to five-fold larger than the values reported in earlier DSS works~\cite{Huang2021_SweetSpots,Cheng2022_Coherence}. The corresponding energy relaxation time at Working Point 3 is $T_1=553\ \mu\mathrm{s}$, which also exceeds the static sweet spot value $T_1=430\ \mu\mathrm{s}$. We summarize these results in Table~\ref{Time_table} and illustrate the periodic modulation used at Working Point 2 as an example in Fig.~\ref{Result_PF}(b). 
{
To further visualize the local robustness of this operating point in experimentally tunable parameters, we perform a two-dimensional scan around Working Point~2, in the physical amplitude--frequency plane \((\tilde{\phi}_{\rm ac},\omega_d)\). 
The resulting \(T_1\) and \(T_\phi\) landscapes are shown in Supplementary Fig.~S2. In particular, the high-\(T_\phi\) ridge indicates that the enhanced dephasing time is not an isolated single-point feature.}

In simulations, we truncate infinite-dimensional objects to finite size, such as the Fourier-coefficient set ${p_n}$ and the Floquet matrix. Specifically, the Floquet matrix is truncated to dimension $3n$, which confines the numerical error in the computed Floquet states to below $10^{-10}$ (see FIG. S4 in Supplementary Sec.~II.B).
We tested Fourier truncation orders \(n=1,\dots,5\) and found that \(n=4\) yields the most complete PF for our parameter set. Accordingly, this section presents results for \(n=4\) only, while results for other \(n\) are given in Supplementary Fig.~S3 and Supplementary Sec.~II.A. For each \(p_n\) with $n\ge 1$, the real and imaginary parts are constrained to lie in \([-1,1]\), whereas the zeroth coefficient \(p_0\) is restricted to \([0,1]\). 
{We impose the constraint \(p_0 \ge 0\), because negative values of \(p_0\) can shift the effective dc bias \(\tilde{\phi}_{\mathrm{dc}}=\phi_{\mathrm{dc}}+\phi_{\mathrm{ac}}p_0\) toward \(\pi\), where the system approaches a static sweet spot.} The drive frequency is set in $\omega_d\in[0.5\Omega_{ge},1.5\Omega_{ge}]$. The device parameters are $E_C/(2\pi)=1$\,GHz, $E_L/(2\pi)=0.79$\,GHz, and $E_J/(2\pi)=4.43$\,GHz. The noise parameters are $A_d=\pi^2\tan(\delta_C)\lvert\tilde{\varphi}_{01}\rvert^2\hbar/E_C$ and $A_f=2\pi\delta_f E_L\lvert\tilde{\varphi}_{01}\rvert$, with $\tan(\delta_C)=1.1\times10^{-6}$, $\delta_f=1.8\times10^{-6}$, and temperature $\mathcal{T}=15$\,mK.

{
We further test the robustness of the Pareto-front results against multilevel corrections of the full fluxonium Hamiltonian. 
As shown in Supplementary Sec.~III.A and Fig.~S5, the quasienergy gap \(\Omega\) and the DSS condition \(g_z^{[0]}\sim \partial\Omega/\partial\tilde{\phi}_{\rm dc}\) converge with increasing level truncation, and the corresponding DSS locations become stable for \(N\ge 5\).
We therefore repeat the PF optimization using an \(N=5\) truncation which is shown in Supplementary Sec.~III.B.
The resulting aggregated PF is nearly unchanged in the objective space compared with the PF obtained from the \(N=2\) truncation, showing that higher fluxonium levels primarily lead to quantitative shifts in the optimized waveform parameters and DSS locations, while preserving the achievable \(T_1\)--\(T_\phi\) trade-off. Representative DSS points selected from the \(N=5\) PF also remain stable under higher truncations with \(N>5\). These results indicate that multilevel effects modify the decoherence rates quantitatively but do not change the main qualitative conclusions of the Pareto-front engineering framework.
}

\subsection{Identifying DSS and the \(T_1\) upper bound}

\begin{table}[t]
    \centering
    \setlength{\tabcolsep}{14pt}
    \begin{tabular}{ccc}
        \hline\hline                                                                     \\[-8pt]
        \textbf{Working points} & \textbf{$T_1$ ($\mu$s)} & \textbf{$T_{\phi}$ ($\mu$s)} \\
        \hline                                                                           \\[-8pt]
        Off static sweet spot   & 940                     & 1                            \\
        DSS-1                   & 877                     & 2036                         \\
        DSS-2                   & 711                     & 3035                         \\
        DSS-3                   & 553                     & 7398                         \\
        Static sweet spot       & 430                     & $>10^4$                     \\
        DSS in Ref.~\cite{Huang2021_SweetSpots} & 590                     & 1200                         \\
        DSS in Ref.~\cite{Cheng2022_Coherence}  & 718                     & 2339                         \\
        \hline\hline
    \end{tabular}
    \caption{
    Coherence times $T_1$ and $T_\phi$ for representative operating points.
    We label three PF points discussed in Fig.~\ref{Result_PF} as DSS-1--DSS-3.
    “Static sweet spot” refers to \(\tilde{\phi}_{\mathrm{dc}}=\pi\), where first-order sensitivity to DC flux noise vanishes and \(T_\phi\to\infty\). “Off static sweet spot” uses \(\tilde{\phi}_{\mathrm{dc}}=1.03\pi\).
    }
    \label{Time_table}
\end{table}

A DSS is an operating point where the quasienergy gap is first-order insensitive to the DC flux, i.e., \(\partial\Omega/\partial\tilde{\phi}_{\mathrm{dc}}\approx 0\). In the Floquet description, this condition can be identified through the zeroth Fourier component \(g_z^{[0]}\). Perturbation theory shows that \(\partial\Omega/\partial\tilde{\phi}_{\mathrm{dc}}\) scales with \(|g_z^{[0]}|\)~\cite{Huang2021_SweetSpots}. Therefore, small \(|g_z^{[0]}|\) provides a direct numerical criterion for identifying DSSs.

In our multi-objective periodic-flux modulation framework, the optimization minimizes both relaxation and dephasing rates. Since the dephasing objective \(\gamma_z\) depends on the coefficient \(g_z^{[0]}\) in Eq.~\eqref{gamma}, the optimization naturally favors working points with suppressed \(|g_z^{[0]}|\). Our numerical results show that all working points to the right of Working Point 1 in Fig.~\ref{Result_PF}(a) satisfy \(|g_z^{[0]}|<10^{-4}\). Using this threshold, this entire region can be regarded as a DSS regime. The three representative points DSS-1--DSS-3 in Table~\ref{Time_table} are selected from this regime. 
{
In this optimization, tuning \(\tilde{\phi}_{\rm dc}\) should be understood as part of selecting the physical operating point, and it does not conflict with the role of the DSS. 
The DSS condition provides local first-order robustness around the selected point after it has been set. It does not eliminate the need for the initial calibration of the physical DC bias required to reach the desired DSS point.
}
{Fig.~\ref{Result_PF}(a) also shows that \(T_1\) does not increase without bound in the DSS region. This behavior can be understood from the long-time relaxation rate. As shown in Supplementary Sec.~IV.A, the total relaxation rate is
\(\gamma_1=\sum_{k\in\mathbb Z}|g_+^{[k]}|^2\tilde S(k\omega_d-\Omega),\)
where \(\tilde S(\omega)\equiv S(\omega)+S(-\omega)\) is the symmetrized noise spectrum. Thus, a finite upper bound on \(T_1=1/\gamma_1\) follows when both the sampled noise spectrum and the total transition-filter weight are bounded from below. For the \(1/f\) flux-noise plus dielectric-loss spectrum in Eq.~\eqref{S}, \(\tilde S(\omega)\) has a positive lower bound. In the DSS regime, the condition \(g_z^{[0]}\approx0\) also imposes a lower bound on \(\sum_k|g_+^{[k]}|^2\) for arbitrary periodic waveforms. This DSS condition is essential: away from the DSS, the corresponding lower bound on \(\sum_k|g_+^{[k]}|^2\) acquires a factor \(1-|g_z^{[0]}|^2\), which becomes zero as \(|g_z^{[0]}|\to1\). Combining the \(\tilde S(\omega)\) lower bound and the lower bound of \(\sum_k|g_+^{[k]}|^2\) gives an upper bound for \(T_1\), which is summarized in the following theorem (with detailed proof in Supplementary Sec.~IV.A).}

{\begin{theorem}\label{th}
Within the effective two-level model, consider the system--bath interaction \(H_{\mathrm{int}}=\sigma_z\otimes\beta\), with the bath noise spectrum described by the \(1/f\) flux-noise plus dielectric-loss model in Eq.~\eqref{S}. If the system is operated in the ideal DSS limit \(g_z^{[0]}=0\), then the energy relaxation time satisfies
\begin{equation}\label{ub_global}
    T_1\leq     \frac{\omega_d^2+\Delta^2}
    {3\omega_d^2(A_f^4A_d\hbar^2)^{1/3}} .
\end{equation}
\end{theorem}
Theorem~\ref{th} gives a waveform-independent upper bound within the DSS regime for the specified \(\sigma_z\)-coupled noise model. It should not be interpreted as a bound that is independent of the noise model itself. For modified spectra that still couple through \(\sigma_z\), the same proof applies as long as the resulting total symmetrized spectrum retains a positive global lower bound. Examples include spectra with ohmic components or with non-\(1/f\) power-law exponents. In such cases, the spectral constant in Eq.~\eqref{ub_global} should be replaced by the corresponding lower-bound constant of the modified spectrum; see Supplementary Sec.~IV.B for details. By contrast, mechanisms involving additional system operators or separate bath channels, such as quasiparticle-induced loss, lie outside the microscopic assumptions of Theorem~\ref{th}. If such channels contribute independently and additively, they increase the total relaxation rate and therefore reduce the observed \(T_1\). 
Thus, the upper bound derived here may overestimate the maximum \(T_1\) achievable in a real device. 
Colored technical noise near the drive frequency \(\omega_d\), such as source phase noise, amplitude noise, spurs, or control-line transfer-function distortions, is also not generally captured by the stationary bath spectrum assumed here and requires a separate control-noise analysis. }

{
\subsection{Construction and robustness of double-DSS}
\label{subsec:double_dss_construction}

We now construct working points that are first-order insensitive to both the physical DC bias and the AC modulation amplitude. In the physical parameter plane \((\tilde{\varphi}_{dc},\tilde{\varphi}_{ac})\), these double-DSS points can be viewed as intersections of two zero-value curves,
 \(F_{dc}(\tilde{\varphi}_{dc},\tilde{\varphi}_{ac})\equiv \partial\Omega/\partial\tilde{\varphi}_{dc}=0\) and \(F_{ac}(\tilde{\varphi}_{dc},\tilde{\varphi}_{ac})\equiv\partial\Omega/\partial\tilde{\varphi}_{ac}=0\)
where \(\Omega\) is the Floquet quasienergy gap. The first condition defines a DSS with respect to DC flux noise, whereas the second condition suppresses first-order sensitivity to fluctuations in the AC modulation amplitude.

Since a double-DSS must also satisfy the usual DSS condition, the DSS points on the PF, which already lie close to \(F_{dc}=0\), provide natural starting points for constructing double-DSS solutions. Starting from a selected PF point, we keep its normalized waveform coefficients \(\{p_k\}_{k\neq0}\) and drive frequency \(\omega_d\) fixed, and vary only the two physical parameters \(\tilde{\varphi}_{dc}\) and \(\tilde{\varphi}_{ac}\). This reduces the construction to a search in the physical parameter plane rather than in the full high-dimensional waveform space. 
The key idea is to first use \(F_{dc}=0\) to restrict the search to a local DSS curve, and then search for \(F_{ac}=0\) along that curve. 
If \(\partial F_{dc}/\partial\tilde{\varphi}_{dc}\neq0\) near the selected PF point, the implicit function theorem guarantees that the DSS curve can be locally parameterized as \(\tilde{\varphi}_{dc}=f(\tilde{\varphi}_{ac})\).
The second condition \(F_{ac}=0\) therefore does not need to be searched over the full two-dimensional plane. Instead, it is sufficient to solve the one-dimensional root-finding problem $G(\tilde{\varphi}_{ac})=F_{ac}\bigl(f(\tilde{\varphi}_{ac}),\tilde{\varphi}_{ac}\bigr)=0$, where \(G\) denotes the value of \(F_{ac}\) restricted to the local DSS curve \(F_{dc}=0\).

\begin{figure}[t]
    \centering
    \includegraphics[width=0.5\textwidth]{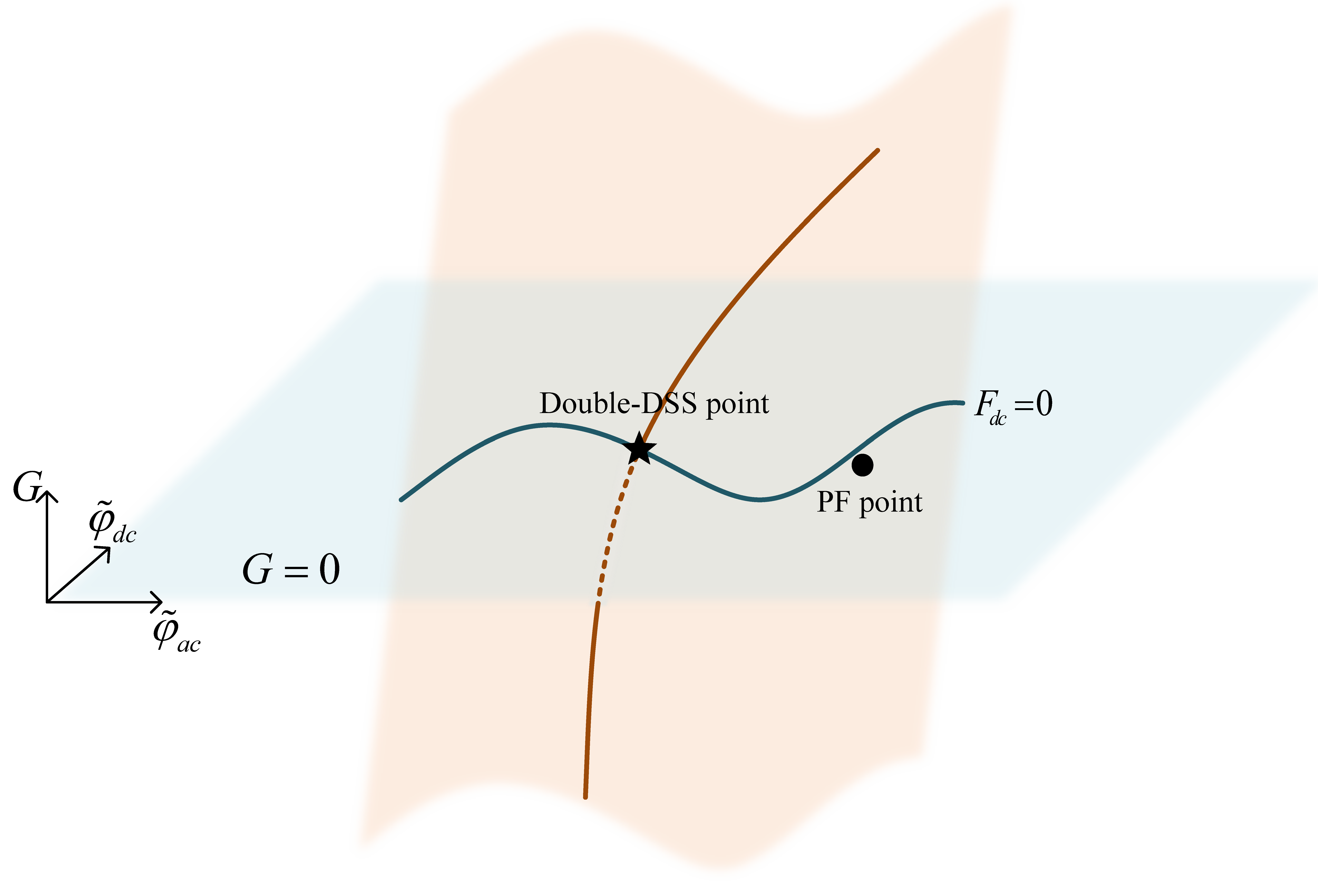}
    \caption{
    Geometric illustration of a double-DSS point. 
    The bottom plane represents the physical parameter plane 
    \((\tilde{\varphi}_{dc},\tilde{\varphi}_{ac})\). 
    The blue curve denotes the local DSS curve defined by \(F_{dc}=0\). 
    Along this curve, we evaluate 
    \(G(\tilde{\varphi}_{ac})=F_{ac}(f(\tilde{\varphi}_{ac}),\tilde{\varphi}_{ac})\), 
    namely the value of \(F_{ac}\) restricted to the DSS curve, and use \(G\) as the vertical coordinate.
    The double-DSS point is identified as the intersection between this lifted curve and the plane \(G=0\). 
    Therefore, its projection lies on the DSS curve and satisfies \(F_{dc}=0\), while its vertical coordinate vanishes and satisfies \(F_{ac}=0\).
    }
    \label{fig:double_dss_geometry}
\end{figure}

This reduction from a two-dimensional intersection problem to a constrained one-dimensional root search is illustrated geometrically in Fig.~\ref{fig:double_dss_geometry}. The bottom plane is the physical parameter plane \((\tilde{\varphi}_{dc},\tilde{\varphi}_{ac})\), where the blue curve denotes the DSS curve \(F_{dc}=0\). Along this curve, we evaluate \(G(\tilde{\varphi}_{ac})\) and use \(G\) as the vertical coordinate, giving a spatial curve on the surface generated by lifting the DSS curve along the \(G\) direction. A double-DSS point corresponds to the intersection between this spatial curve and the plane \(G=0\): its projection lies on the DSS curve and therefore satisfies \(F_{dc}=0\), while its vertical coordinate vanishes and therefore satisfies \(F_{ac}=0\).

Numerically, we search for double-DSS points within the preselected local physical window: \(\tilde{\Phi}_{dc}/\Phi_0\in[0.521,0.523]\) and \(\tilde{\Phi}_{ac}/\Phi_0\in[0.070,0.080]\). This window lies on the right side of the half-flux point and corresponds to a moderate flux modulation amplitude that is experimentally relevant. It therefore provides a local and physically accessible working region, rather than an unconstrained search over the full high-dimensional waveform space. This local window is not intended to exhaust all possible double-DSS solutions; a broader physical search window can in principle yield additional double-DSS points. Within this window, we discretize \(\tilde{\varphi}_{ac}\). For each fixed value of \(\tilde{\varphi}_{ac}\), we solve \(F_{dc}=0\) along the \(\tilde{\varphi}_{dc}\) direction to obtain a sampled point on the local DSS curve. We then evaluate \(G=F_{ac}\) only on these sampled DSS points. When \(G\) changes sign between two neighboring samples, we solve \(G=0\) along the DSS curve to obtain a double-DSS point satisfying both \(F_{dc}=0\) and \(F_{ac}=0\). Details of the discretization, sign-change detection, and root-finding procedure are provided in Supplementary Sec.~V.A.

Within this local window, we obtain 29 double-DSS points. To assess their benefit, we introduce a pure-dephasing model that includes both DC flux noise and AC amplitude noise, and denote the corresponding effective pure-dephasing time by \(T_{\phi,dc+ac}=1/\gamma_{\phi}^{dc+ac}\). Here \(\gamma_{\phi}^{dc+ac}=\gamma_{\phi}^{dc}+\gamma_{\phi}^{ac}\), where \(\gamma_{\phi}^{dc}\) is the pure-dephasing rate from the original DC flux-noise channel, and \(\gamma_{\phi}^{ac}\) is the additional pure-dephasing rate induced by AC amplitude noise. The coupling operator associated with AC amplitude noise is \(V_{ac}(t)=\partial H(t)/\partial\tilde{\varphi}_{ac}=\tilde{P}(t)\sigma_z\), where \(\tilde{P}(t)\) is the normalized drive waveform. Taking the difference between the diagonal matrix elements of \(V_{ac}(t)\) in the two Floquet modes gives the pure-dephasing harmonic coefficients \(g_{z,ac}^{[k]}\) for the AC-noise channel. The resulting \(\gamma_{\phi}^{ac}\) can be written as
\begin{equation}
    \gamma_{\phi}^{ac}=|\partial\Omega/\partial\tilde{\varphi}_{ac}|A_{f,ac}\sqrt{2|\ln(\omega_{ir}t_m)|}+\sum_{k\neq0}|g_{z,ac}^{[k]}|^2S_{ac}(k\omega_d).
\end{equation}
Here \(A_{f,ac}\) denotes the low-frequency \(1/f\) noise amplitude associated with fluctuations of the AC drive amplitude. In the numerical comparison below, we take \(A_{f,ac}=A_f\), so that the DC-flux and AC-amplitude noise channels are compared on the same low-frequency noise scale. This derivation is given in Supplementary Sec.~V.B. 

For the 29 double-DSS points, the pure-dephasing time under the original dc-only dephasing model is \(T_{\phi}^{\mathrm{double}}\in[4443.043,5899.603]\ \mu\mathrm{s}\) as shown in Fig.~\ref{Result_PF}(a). After including the AC amplitude-noise channel, the effective pure-dephasing time becomes \(T_{\phi,dc+ac}^{\mathrm{double}}\in[4440.313,5897.479]\ \mu\mathrm{s}\). Thus, the double-DSS points remain nearly unaffected by the additional AC-amplitude dephasing channel, as expected from the condition \(\partial\Omega/\partial\tilde{\varphi}_{ac}=0\). To evaluate their operational relevance, we compare these double-DSS points with single-DSS points at the same or closest \(T_1\). The matched single-DSS points have \(T_{\phi,dc+ac}^{\mathrm{single}}\in[25.480,25.931]\ \mu\mathrm{s}\). Therefore, the improvement ratio of the double-DSS points over the matched single-DSS points is \(T_{\phi,dc+ac}^{\mathrm{double}}/T_{\phi,dc+ac}^{\mathrm{single}}\in[174.2,227.4]\). Based on the median, the double-DSS points provide a 215-fold enhancement in \(T_{\phi,dc+ac}\). These results show that double-DSS points are not merely geometric intersections of two zero-sensitivity curves; even when both DC and AC pure-dephasing channels are included, they still provide substantially longer effective pure-dephasing times than matched single-DSS operating points.
}

\section{Realization of quantum gates}\label{sec:gates}

{
In the previous sections, we have shown that DSS modulation can simultaneously improve energy relaxation and dephasing at selected operating points.
We now show that such optimized periodically driven operating points can also support quantum gate operations. Previous work has shown that logical operations can be defined on Floquet states~\cite{Huang2021_SweetSpots}. In this work, the computational space is still spanned by two and four Floquet computational states. However, the physical control pulse acts on the full multilevel Hilbert space and can populate levels outside the computational subspace.  
To account for leakage outside the Floquet computational subspace, we optimize the gate pulses in a multilevel truncation.
Since the multilevel convergence analysis in Supplementary Sec.~III shows that the DSS location becomes stable for \(N\ge 5\), we use the \(N=5\) truncation for pulse training and validate the optimized pulses in larger truncations below. 
Motivated by gate-control protocols for fluxonium qubits~\cite{Nesterov2022_CNOT}, we keep the device at a selected gate working point and implement the target gates by superimposing additional charge-drive control pulses on top of the periodic background. 
The modulation parameters of this point are listed in Supplementary Table~S1. 
In the absence of the additional gate pulses, this point has idle coherence times 
\(T_1^{\rm idle}=872.6~\mu{\rm s}\) and 
\(T_\phi^{\rm idle}=1970.6~\mu{\rm s}\). For the two-qubit gate, both fluxonium qubits are operated at the same gate working point before the local charge-control pulses are applied.}

\subsection{Gate implementation in periodically driven systems}
\label{subsec:gate_framework}

{
For an \(N\)-level truncated model, we omit the superscript \(N\) in this subsection unless the truncation dimension needs to be emphasized explicitly. The total Hamiltonian during the gate is written as
\begin{equation}
H_G(t)=H_q(t)+\sum_\mu f_\mu(t)\hat n_\mu .
\label{eq:gate_total_hamiltonian}
\end{equation}
Here \(H_q(t)\) is the DSS periodic background Hamiltonian within the chosen \(N\)-level truncation, \(\hat n_\mu\) is the projected charge operator for the \(\mu\)-th control channel, and \(f_\mu(t)\) is the physical envelope multiplying this operator. The corresponding logical pulse in the computational subspace is defined below. Such charge control can be implemented with capacitively coupled microwave lines~\cite{Krantz2019_Guide}. We use one local charge-control channel for single-qubit gates and two local channels for two-qubit gates.

Let \(\hat n_{\mu}^{\rm full}\) denote the physical charge operator in the original Hilbert space, and let \(\{|\varphi_j\rangle\}\) be eigenstates of the static sweet-spot Hamiltonian. We take the first \(N\) eigenstates to form the truncated basis
\(
W_N=
\begin{bmatrix}
|\varphi_0\rangle,
|\varphi_1\rangle,
\cdots,
|\varphi_{N-1}\rangle
\end{bmatrix}.
\)
Projecting the full charge operator into this \(N\)-dimensional subspace gives $\hat n_\mu=W_N^\dagger\hat n_{\mu}^{\rm full}W_N .$
Thus, the control operators for different truncation dimensions are all obtained from the same physical charge operator; the only difference is the number of retained eigenstates.

We implement the logical gate on the two Floquet basis states $|\omega_+(0)\rangle$ and $|\omega_-(0)\rangle$ within this \(N\)-level truncation. Let \(U_q(t)\) be the evolution generated only by the periodic background Hamiltonian, satisfying $i\dot U_q(t)=H_q(t)U_q(t).$ 
In the rotating frame defined by \(U_q(t)\), the charge-control operator becomes $\widetilde n_\mu(t)=U_q^\dagger(t)\hat n_\mu U_q(t).$
Let \(U_{\rm tot}(t)\) be the total evolution generated by Eq.~\eqref{eq:gate_total_hamiltonian} and \(T_g\) denote the total gate duration. The rotating-frame propagator at the end of the gate is then defined as
\(\widetilde U(T_g)=U_q^\dagger(T_g)U_{\rm tot}(T_g)\).
In the single-qubit case, we define the matrix \(F\) as the encoding matrix from the two-dimensional logical space into the \(N\)-level space,
$F=\begin{bmatrix}
|\omega_+(0)\rangle,
|\omega_-(0)\rangle
\end{bmatrix}
\in\mathbb C^{N\times2}.$ For any $|\psi\rangle\in \mathcal{H}^2$, one has $F|\psi\rangle \in \mathcal{H}^{N}$.
For the two-qubit system, the corresponding matrix is $F=F_L\otimes F_R
\in \mathbb C^{N^2\times4}.$
Given the actual multilevel rotating-frame propagator \(\widetilde U(T_g)\), the matrix
\(F^\dagger \widetilde U(T_g)F\)
is the effective logical operator of the actual \(N\)-level evolution in the Floquet logical basis.
Let \(U_d\) denote the target gate in the Floquet logical basis. This matrix is specified in the abstract \(d\)-dimensional logical coordinates and does not depend on the truncation dimension \(N\). We therefore define the error matrix as
$M=U_d^\dagger F^\dagger \widetilde U(T_g) F .$
If the implemented gate equals the target gate, then \(M=I_d\). We quantify the gate quality using the average fidelity~\cite{PEDERSEN200747}:
\begin{equation}
F_{\rm avg}=\frac{{\rm Tr}(MM^\dagger)+|{\rm Tr}(M)|^2}{d(d+1)}.
\label{eq:gate_average_fidelity}
\end{equation}
Here \(d\) is the computational-subspace dimension, with \(d=2\) for a single qubit and \(d=4\) for two qubits.
Throughout this section, \(F_{\rm avg}\) denotes the average gate fidelity including leakage contributions. 
With \(P_L=I-FF^\dagger\), the average leakage is defined as $L_{\rm avg}=\frac{1}{d}\left\|P_L\widetilde U(T_g)F\right\|^2 $, which can be also written as $L_{\rm avg}=1-\frac{1}{d}{\rm Tr}(M^\dagger M).$
This average leakage measures the average population that flows into the non-computational subspace after the full \(N\)-level evolution, when the initial state is in the Floquet computational subspace.

In the pulse optimization, we minimize
\begin{equation}
\mathcal L=1-F_{\rm avg}+\lambda_{\rm ch}\mathcal L_{\rm ch}.
\label{eq:gate_loss_function}
\end{equation}
Here \(\mathcal L_{\rm ch}\) suppresses first-order leakage from the Floquet computational subspace to selected high-level non-computational states.
It is inspired by the spectral-leakage suppression principle
underlying DRAG and FAST-DRAG pulse design
\cite{Motzoi2009DRAG,Chow2010DRAG,Hyyppa2024FASTDRAG}, but is formulated
directly in the Floquet rotating frame.
The parameter \(\lambda_{\rm ch}\) sets its weight in the total loss.
Let \(P_{\rm ch}\) denote the projector onto those selected states. The first-order Dyson expansion gives the corresponding leakage-amplitude matrix
\(A_{\rm ch}^{(1)}=-i\sum_\mu\int_0^{T_g} f_\mu(t)P_{\rm ch}\widetilde n_\mu(t)F\,dt\),
and we define \(\mathcal L_{\rm ch}=\frac{1}{d}\|A_{\rm ch}^{(1)}\|_F^2\), where $\|\cdot\|_F$ is Frobenius norm. Since this term is evaluated from the first-order leakage amplitude rather than an additional full propagation, it provides a low-cost way to guide the search away from dominant leakage channels. 
The specific high-level states included in \(P_{\rm ch}\) are specified in the corresponding gate subsections. To facilitate the numerical implementation, we provide the frequency-domain form of \(\mathcal L_{\rm ch}\) in Supplementary Sec.~VI.A. 

To compute \(\widetilde U(T_g)\), we discretize the gate interval \([0,T_g]\) into uniform time steps. Under a first-order Lie--Trotter approximation,
\begin{equation}
\widetilde U(T_g)\approx\prod_{k=1}^{N_t}\exp\left[-i\Delta t\sum_\mu f_\mu(t_k)\widetilde n_\mu(t_k)\right].
\label{eq:grape_propagator_general}
\end{equation}
The discrete control amplitudes are updated using a GRAPE-type optimization algorithm. To ensure experimental feasibility, we use a bandwidth-limited frequency-domain parameterization and impose endpoint, amplitude, and bandwidth constraints~\cite{Song2022_Autodiff}.

\begin{figure*}[tb]
    \centering
    \includegraphics[width=\linewidth]{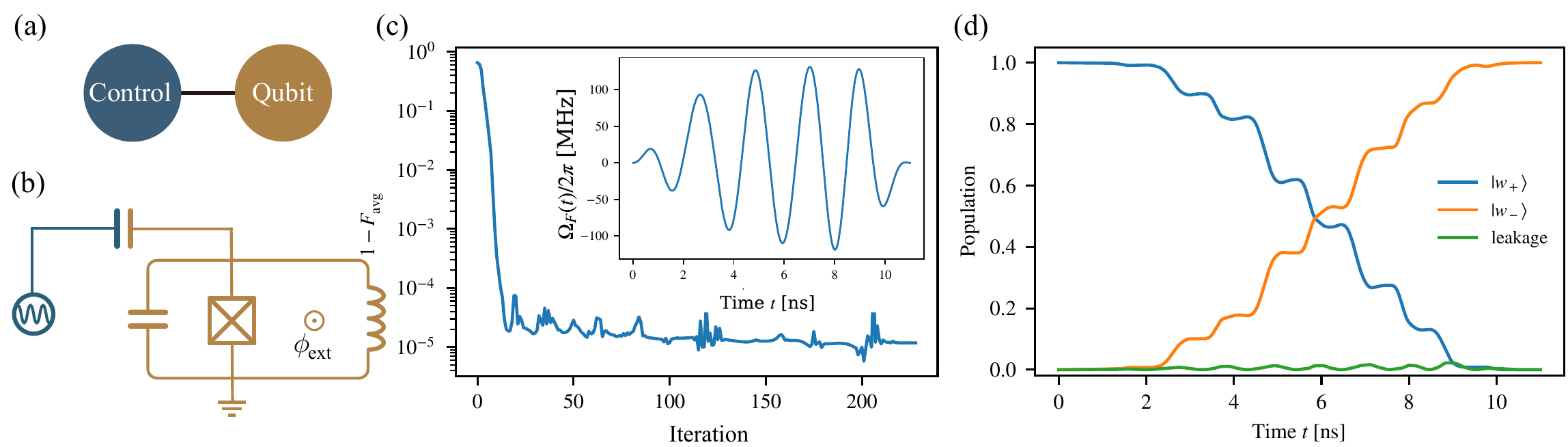}
    \caption{
    (a) Control setup for implementing a single-qubit gate in the fluxonium system.
    (b) Device circuit. The system is periodically modulated at the gate working point, while an additional charge-drive control pulse is applied to implement the gate.
    (c) Closed-system infidelity \(1-F_{\rm avg}\) versus GRAPE iteration for the single-qubit \(X\) gate. The inset shows the optimized pulse \(\Omega_F(t)\). The total gate duration is \(11~{\rm ns}\), and the pulse amplitude satisfies \(\max|\Omega_F|/2\pi=110~{\rm MHz}\).
    (d) Population dynamics for the representative initial state \(|\omega_+(0)\rangle\), including the two Floquet computational-state populations and the leakage population outside the computational subspace.
    }
    \label{X_gate}
\end{figure*}

The Rabi-control amplitude in a conventional two-level model is an effective control amplitude defined after projecting the physical control operator into the logical subspace. We use the same idea here, but with the Floquet computational subspace. For a physical control envelope \(f_\mu(t)\), the corresponding contribution to the logical Hamiltonian is \(f_\mu(t)C_\mu(t)\), where
\(C_\mu(t)=F^\dagger\widetilde n_\mu(t)F\)
is the \(d\times d\) representation of the rotating-frame charge operator in the Floquet computational basis. 
In this basis, \(C_\mu(t)\) can be separated into an identity component and a traceless component. The identity component only produces a global phase and does not generate a logical rotation. We therefore use the operator norm of the traceless part to define the Floquet-logical control scale:
\begin{equation}
\eta=\max_{\mu,k}\left\|C_\mu(t_k)-\frac{{\rm Tr}[C_\mu(t_k)]}{d}I_d\right\|_{\rm op}.
\label{eq:floquet_control_scale}
\end{equation}
Thus, \(\eta\) is the largest nontrivial logical control strength generated by a physical envelope, over all sampled times and control channels.  For the single- and two-qubit optimizations considered below, this definition gives the same value, \(\eta=0.105\). We then define the Floquet-logical pulse amplitude as \(\Omega_{F,\mu}(t)=\eta f_\mu(t)\). In the following pulse optimization and figures, we constrain and display \(\Omega_{F,\mu}(t)\), while the actual multilevel propagation still uses the physical control envelope \(f_\mu(t)\).}

\subsection{Single-Qubit Gate}
\label{subsec:single_qubit_X_gate}

{
We first implement a single-qubit gate by applying an additional control pulse while keeping the device at the gate working point.
The control schematic is shown in Fig.~\ref{X_gate}(a), and the corresponding quantum circuit is shown in Fig.~\ref{X_gate}(b). The control Hamiltonian in the \(N\)-level model is
\begin{equation}
H(t)=H_q(t)+\Omega_{F}(t)\frac{\hat n}{\eta} .
\label{eq:X_gate_hamiltonian}
\end{equation}
As an example, we implement an \(X\) gate with total duration \(T_g=11~{\rm ns}\). The control pulse is parameterized in the frequency domain, keeping only the first 8 Fourier components. During the optimization, we constrain the Floquet-logical pulse amplitude \(\Omega_{F}(t)\) and set \(\max |\Omega_{F}|/2\pi=110~{\rm MHz}\). 
The pulse is trained in an \(N=5\) multilevel model using the objective function in Eq.~\eqref{eq:gate_loss_function}. During training, \(P_{\rm ch}\) is chosen as the projector onto the relevant high-level non-computational states identified from the \(N=6\) and \(N=7\) truncated models, in order to suppress the dominant first-order leakage channels to these levels.
The main panel of Fig.~\ref{X_gate}(c) shows the training infidelity 
\(1-F_{\rm avg}^{(N=5)}\) during the GRAPE optimization in the \(N=5\) multilevel model. 
At the final iteration, the training-model average fidelity reaches 
\(F_{\rm avg}^{(N=5)}=99.99\%\). 
The inset of Fig.~\ref{X_gate}(c) shows the corresponding optimized Floquet-logical control pulse 
\(\Omega_{F}(t)\).

After fixing this optimized pulse, we validate it in a higher-dimensional model with 
\(N_{\rm val}=12\), where \(F_{\rm avg}\) and \(L_{\rm avg}\) are recalculated from the full 
multilevel propagator. 
The validation values are 
\(F_{\rm avg}^{(N_{\rm val}=12)}=99.99\%\) and 
\(L_{\rm avg}^{(N_{\rm val}=12)}=9.35\times 10^{-6}\), respectively.
To visualize the gate action, Fig.~\ref{X_gate}(d) shows the population dynamics for the initial state 
\(|\omega_+(0)\rangle\), including both the computational-subspace populations and the leakage 
population outside the computational subspace. 
The pulse spectrum, the time-dependent average leakage \(L_{\rm avg}(t)\), and the truncation-convergence result are given in Supplementary Sec.~VI.B.
}

\begin{figure*}[t]
    \centering
    \includegraphics[width=\linewidth]{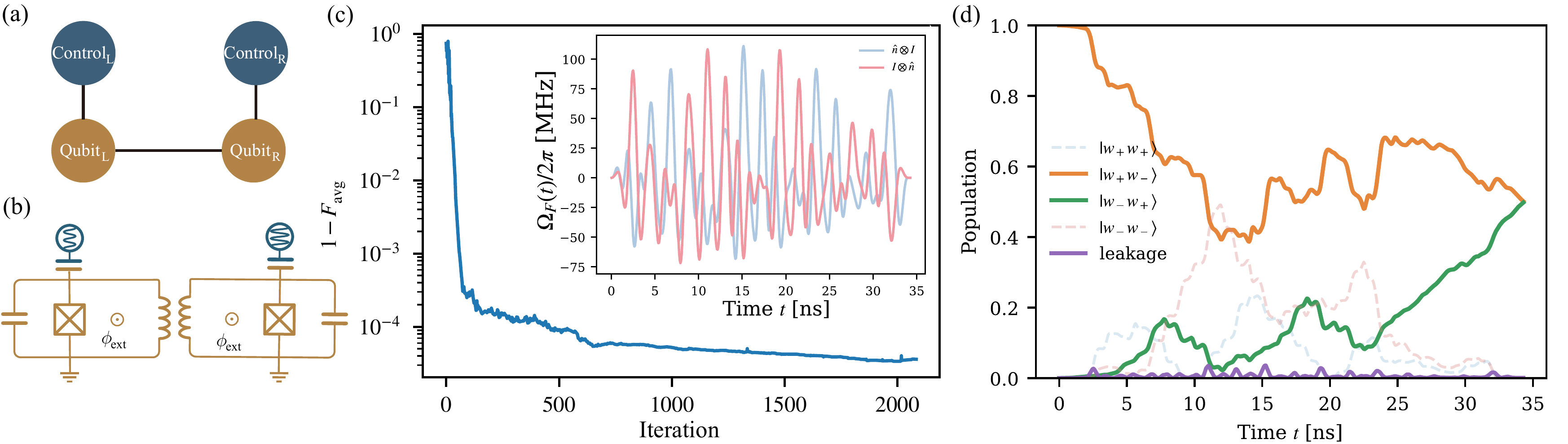}
    \caption{
        (a) Control setup for implementing a two-qubit gate between two fluxonium qubits.
        (b) Device circuit. Both qubits are periodically modulated at the same gate working point and are driven by independent local charge-control pulses. The two-qubit interaction is provided by a fixed flux-flux coupling.
        (c) Closed-system infidelity \(1-F_{\rm avg}\) versus GRAPE iteration for the \(\sqrt{i{\rm SWAP}}\) gate. The inset shows the optimized local pulses \(\Omega_{F,1}(t)\) and \(\Omega_{F,2}(t)\). The total gate duration is \(34~{\rm ns}\), and the amplitudes satisfy \(\max_j|\Omega_{F,j}|/2\pi=110~{\rm MHz}\).
        (d) Population dynamics for the representative initial state \(|\omega_{+-}(0)\rangle\), including the four Floquet computational-state populations and the leakage population outside the computational subspace.
    }
    \label{iswap_gate}
\end{figure*}

\subsection{Two-Qubit Gate}
\label{subsec:two_qubit_iswap_gate}

{
To implement a two-qubit gate, we operate both fluxonium qubits at the gate working point and apply local charge-drive control pulses to each qubit.
The control schematic is shown in Fig.~\ref{iswap_gate}(a), and the corresponding quantum circuit is shown in Fig.~\ref{iswap_gate}(b). The total Hamiltonian for the two-qubit gate in the \(N\)-level model is
\begin{equation}
H(t)=H_{2q}(t)+H_{\rm int}+\Omega_{F,L}(t)\frac{\hat n_L}{\eta}+\Omega_{F,R}(t)\frac{\hat n_R}{\eta} .
\label{eq:two_qubit_hamiltonian}
\end{equation}
Here \(H_{2q}(t)\) contains the DSS background Hamiltonians of the two single qubits. The operators \(\hat n_L=\hat n_L\otimes I_R\) and \(\hat n_R=I_L\otimes \hat n_R\) denote the local charge-control operators acting on the left and right qubits, respectively. The two-qubit interaction is taken to be a flux-flux coupling: 
\(H_{\rm int}=J_{\rm raw}Q_L\otimes Q_R\), with \(J_{\rm raw}=48\) MHz, where \(Q_\mu=\pi I_\mu-\hat\phi_\mu\) is the flux-displacement operator relative to \(\phi_{\rm ext}=\pi\).
Let \(S_\mu=(|\tilde g_\mu\rangle,|\tilde e_\mu\rangle)\) denote the static lowest-two-level basis of fluxonium qubit \(\mu=L,R\). Since the projected flux-displacement operator is predominantly transverse, \(S_\mu^\dagger Q_\mu S_\mu\propto\tilde{\sigma}_x\), the multilevel coupling reduces in the two-level subspace to an effective \(XX\) interaction, \(H_{\rm int}^{(2)}\simeq J_{\rm eff}\,\tilde{\sigma}_x\otimes\tilde{\sigma}_x\). Here \(J_{\rm eff}\) includes both \(J_{\rm raw}\) and the two flux-displacement matrix elements.

As an example, we implement a \(\sqrt{i{\rm SWAP}}\) gate with total duration \(T_g=34~{\rm ns}\). The two local control pulses \(\Omega_{F,L}(t)\) and \(\Omega_{F,R}(t)\) are parameterized in the frequency domain, each retaining the first 29 Fourier components. During the optimization, we set \(\max |\Omega_{F,j}|/2\pi=110~{\rm MHz}\;(j=L,R)\). The two pulses are trained in an \(N=5\) multilevel model using the same objective function in Eq.~\eqref{eq:gate_loss_function}. During training, \(P_{\rm ch}\) includes the relevant high-level non-computational states from the \(N=6,7,8\) truncated models to suppress the dominant first-order leakage channels. Figure~\ref{iswap_gate}(c) summarizes the \(N=5\) training result: the main panel shows the GRAPE convergence of the training infidelity 
\(1-F_{\rm avg}^{(N=5)}\), and the inset shows the optimized local control pulses \(\Omega_{F,L}(t)\) and \(\Omega_{F,R}(t)\). 
At the final iteration, the training-model average fidelity reaches 
\(F_{\rm avg}^{(N=5)}=99.98\%\). 
We then fix these pulses and validate them in the higher-dimensional model with 
\(N_{\rm val}=12\), obtaining 
\(F_{\rm avg}^{(N_{\rm val}=12)}= 99.92\%\) and 
\(L_{\rm avg}^{(N_{\rm val}=12)}=7.18\times 10^{-4}\).

To visualize the gate action, Fig.~\ref{iswap_gate}(d) shows the population dynamics for the representative initial state 
\(|\omega_{+-}(0)\rangle\). 
The final population remains mainly within the target exchange subspace spanned by 
\(|\omega_{+-}(0)\rangle\) and \(|\omega_{-+}(0)\rangle\), while the population outside the computational subspace remains small. 
The pulse spectra, the time-dependent average leakage \(L_{\rm avg}(t)\), and the truncation-convergence result are given in Supplementary Sec.~VI.B.
}

\subsection{Gate fidelity with decoherence rates}
\label{subsec:gate_fidelity_control_modified_decoherence}

{
During a gate operation, the control pulse modifies not only the coherent dynamics but also the local Floquet structure of the system, and hence the decoherence rates. 
Using the gate working point introduced at the beginning of this section as the idle reference, we include a time-dependent decoherence-rate model in the gate simulation.

For the single-qubit \(X\) gate, the optimized physical control envelope \(f(t)\) is generally not periodic. Therefore, the full gate Hamiltonian is not globally periodic, and a global Floquet treatment cannot be applied directly to the full gate evolution. Instead, we define a local reference periodic Hamiltonian at each gate time, from which we extract the instantaneous Floquet basis and the corresponding decoherence rates. 
Specifically, around each gate time \(t_k\), we choose one Floquet-period window $I_k=\left[t_k-\frac{T}{2},\;t_k+\frac{T}{2}\right]$, where $T=\frac{2\pi}{\omega_d}$ is the period of \(H_q(t)\).
We define the period-averaged control amplitude within this window as $\bar f(t_k)=\frac{1}{T}\int_{I_k} f(\tau)\,d\tau .$
Here \(\bar f(t_k)\) is not the pulse actually applied in the gate. It is an auxiliary parameter used only to construct the local Floquet reference model. The actual gate propagation still uses the original control pulse \(f(t)\). 
In the numerical implementation, \(f(t)\) sets to zero outside this interval when evaluating the local average near the boundaries.
With \(\bar f(t_k)\), we define the local reference periodic Hamiltonian on the window \(I_k\) as
\begin{equation}
H_{\rm ref}^{(t_k)}(\tau)=H_q(\tau)+\bar f(t_k)\hat n ,
\label{eq:local_reference_hamiltonian_reply_symbol}
\end{equation}
where \(\tau\in I_k\). Since \(\bar f(t_k)\) is constant with respect to \(\tau\), and \(H_q(\tau)\) is \(T\)-periodic, we have $H_{\rm ref}^{(t_k)}(\tau+T)=H_{\rm ref}^{(t_k)}(\tau).$
Thus, at each \(t_k\), one can define local Floquet states, local quasienergy gaps, and local decoherence rates.
This local construction has an additional advantage: over the window \(I_k\), the reference Hamiltonian and the actual gate Hamiltonian have the same first-order Magnus term,
\begin{align}
\Omega_{1}^{\rm ref}
&=-i\int_{I_k}\left[H_q(\tau)+\bar f(t_k)\hat n\right]d\tau \\
&=-i\int_{I_k}H_q(\tau)\,d\tau-i\left(\int_{I_k}f(\tau)\,d\tau\right)\hat n
=\Omega_{1}^{\rm gate}.\nonumber
\label{eq:magnus_reference}
\end{align}
In the second equality, we used \(\bar f(t_k)T=\int_{I_k}f(s)\,ds\).
At each \(t_k\), we solve the local Floquet problem defined by \(H_{\rm ref}^{(t_k)}(\tau)\), and insert the resulting local quasienergy gaps and Floquet states into the Floquet decoherence-rate formulas derived in Eq.\eqref{gamma}. This gives the time-dependent rates \(\gamma_\pm(t_k)\) and \(\gamma_\phi(t_k)\), with \(\gamma_1(t_k)=\gamma_+(t_k)+\gamma_-(t_k)\). The same local Floquet states are also used to construct the instantaneous jump operators \(\widetilde\sigma_z(t_k)\) and \(\widetilde\sigma_-(t_k)\).
In the propagation, we take the continuum limit \(t_k\rightarrow t\), obtaining time-dependent quantities $\gamma_1(t),\;\gamma_\phi(t),\;\widetilde\sigma_-(t),\;\widetilde\sigma_z(t).$
The single-qubit open-system gate evolution is then described by
\begin{equation}
\begin{split}
    \frac{d\widetilde{\rho}(t)}{dt}
    =&-i\left[\widetilde H(t),\widetilde\rho(t)\right]
    +\gamma_1(t)\mathcal D[\widetilde\sigma_-(t)]\widetilde\rho(t)\\
    &+\gamma_\phi(t)\mathcal D[\widetilde\sigma_z(t)]\widetilde\rho(t),
\end{split}
\label{eq:single_gate_td_lindblad_reply_symbol}
\end{equation}
where $\mathcal D[L]\rho=L\rho L^\dagger-\frac{1}{2}\left\{L^\dagger L,\rho\right\}.$
The coherent term in Eq.~\eqref{eq:single_gate_td_lindblad_reply_symbol} is still generated by the actual gate pulse \(f(t)\). The period-averaged amplitude \(\bar f(t)\) is used only to compute the local Floquet decoherence rates and jump operators.

After the Lindblad propagation, the evolution defines a quantum channel \(\mathcal E_T\) on the full truncated Hilbert space. We evaluate its action on the Floquet computational subspace by projection,
\(\mathcal E_F(\rho)=F^\dagger\mathcal E_T\left(F\rho F^\dagger \right)F .\)
The open-system average gate fidelity is then computed from the Choi overlap between the projected logical channel and the target unitary channel,
\begin{equation}
    F_{\rm avg}^{\rm open}
    =\frac{{\rm Tr}\left[J_{U_d}^\dagger J_{\mathcal E_F}\right]+d}
    {d(d+1)},
\end{equation}
where \(J_{\mathcal E_F}\) and \(J_{U_d}\) are the unnormalized Choi matrices of \(\mathcal E_F\) and the target unitary channel, respectively. 
As shown above, the coherent leakage of the optimized gates ranges from \(10^{-6}\) for the single-qubit gate to \(10^{-4}\) for the two-qubit gate.
Its contribution to the present open-system infidelity is therefore not dominant, and we neglect the small non-trace-preserving correction induced by leakage in the projected logical channel. With this time-dependent decoherence-rate model, the single-qubit open-system average gate fidelity is \(F_{\rm avg}^{\rm open}=99.99\%\) for both the \(N=5\) training truncation and the \(N_{\rm val}=12\) validation truncation.

For the two-qubit gate, we use the same local-reference Floquet construction. The coherent dynamics is still generated by the two-qubit Hamiltonian in Eq.~\eqref{eq:two_qubit_hamiltonian}, including the fixed flux-flux coupling and the two local charge-control pulses. For the dissipative part, we construct a local reference Hamiltonian separately for each qubit using its local physical control envelope,
\(H_{{\rm ref},\mu}^{(t_k)}(\tau)=H_{q,\mu}(\tau)+\bar f_\mu(t_k)\hat n_\mu,\)
where \(\bar f_\mu(t_k)=\frac{1}{T}\int_{I_k}f_\mu(s)\,ds ,\) and \(\mu=L,R\). This gives \(\gamma_{1,\mu}(t)\), \(\gamma_{\phi,\mu}(t)\), and the corresponding local jump operators. In the two-qubit Lindblad equation, the coherent part uses the full \(\widetilde H_{2q}(t)\), while the dissipative part is the sum of the local dissipative channels for the left and right qubits. With this time-dependent decoherence-rate model, the two-qubit open-system average gate fidelities are
\(99.96\%\) and \(99.90\%\) in the \(N=5\) and \(N_{\rm val}=12\) truncations, respectively.
}

\section{Conclusion}\label{sec:conclusion}

In this work we developed a flux-modulation framework based on general periodic modulation and introduced a Pareto-front optimization strategy to quantify the trade-off between $T_1$ and $T_\phi$ in the presence of low-frequency flux noise and dielectric loss.
The resulting PF identifies non-dominated operating points and makes explicit the optimal trade-off between relaxation and pure dephasing under periodic flux modulation.
{Within the effective two-level Floquet model and the specified $\sigma_z$-coupled noise spectrum, we further derived an upper bound on the achievable $T_1$ in the DSS regime, showing that relaxation cannot be suppressed without limit by waveform engineering alone.}
{Using the optimized PF, we also constructed double-DSS points that retain much longer effective $T_\phi$ than matched single-DSS points at comparable $T_1$ under combined DC-flux and AC-amplitude dephasing.}
{To test the robustness of these PF conclusions beyond the effective two-level approximation, we verified with a five-level fluxonium truncation that higher levels lead to quantitative shifts of the DSS locations and decoherence rates, while leaving the PF structure in the objective space qualitatively unchanged.}

At representative optimized operating points, we design additional charge-drive control pulses to realize quantum gates without moving the device away from the DSS. 
{The gates are optimized and validated in a multilevel Floquet rotating frame, where leakage outside the Floquet computational subspace is explicitly included. The resulting single- and two-qubit gates maintain high average fidelities while keeping multilevel leakage strongly suppressed, and the corresponding open-system simulations remain consistent with high-fidelity operation.}

These results demonstrate the feasibility of realizing high-coherence quantum control through dynamical modulation. 
The proposed computational framework is general and extensible, allowing the incorporation of system-specific noise models in future studies. 
{In particular, quasiparticle generation and quasiparticle-induced decay and dephasing under driven operation are not captured by the present \(\sigma_z\)-coupled noise model and should be incorporated in future DSS gate-error analyses.}
On the other hand, experimental considerations---such as device anharmonicity, control bandwidth and others---may influence optimal designs. In such cases, preference-based multi-objective optimization\cite{CoelloZ00,li2023interactive} can be employed to select PF operating points tailored to specific experimental goals. Moreover, the Pareto-front engineering techniques developed here are broadly applicable and can be extended to other superconducting platforms, including transmon qubits.

\section*{Data availability}
The data supporting the findings of this study are available from the corresponding author, Xiaoting Wang, upon request.

\section*{Code availability}
The code used in this study is available from the corresponding author, Xiaoting Wang, upon request.

\section*{Acknowledgements}
This work was supported by the National Natural Science Foundation of China (Grant No.~92265208, 62173201, U2441217, 92565107), the National Key Research and Development Program of China (Grant No.~2022YFA1405900), and the Innovation Fund of Aerospace Institute 771 (Grant No. 771CX2022003). 
X.-H.~Deng acknowledges support from the Shenzhen Science and Technology Program (KQTD20200820113010023) and the Key-Area Research and Development Program of Guangdong Province (Grant No.~2018B030326001). 
G.~W.~Deng acknowledges support from the Sichuan Science and Technology Program (Grant No.~2024YFHZ0372).

\section*{Author contributions}

Z.~Y. performed the main research, derived the proof, wrote the initial draft and prepared all figures. S.~J. and Y.~H. validated the theory numerically. R.-B.~W. proposed the optimization strategy and feasible control designs. Z.~Y., X.~W., X-.H.~D., R.-B.~W and G.~D. conducted the model analysis. X.~W. supervised the project. All authors contributed to manuscript revision and approved the final version.

\nocite{*}
\bibliography{ref}

\clearpage
\newpage
\onecolumngrid
\vspace*{1cm}

\begin{center}
    \textbf{\large Supplemental Material for \\[5pt] Pareto Front Engineering of Dynamical Sweet Spots in Superconducting Qubits}
\end{center}

\setcounter{equation}{0}
\setcounter{figure}{0}
\setcounter{table}{0}
\setcounter{section}{0}
\makeatletter
\renewcommand{\theequation}{S\arabic{equation}}
\renewcommand{\thefigure}{S\arabic{figure}}
\renewcommand{\bibnumfmt}[1]{[S#1]}
\renewcommand{\citenumfont}[1]{S#1}
\renewcommand{\thesection}{\Roman{section}}
\makeatother

{
\section{Two-level projection and coupling convention}
\label{sec:two_level_projection}

\subsection{Two-level Hamiltonian and Pauli convention}

We provide the two-level projection leading to Eq.~(3) of the main text and clarify the Pauli convention used throughout the manuscript. Starting from the full fluxonium Hamiltonian in Eq.~(1) of the main text,
\begin{equation}
H_{\mathrm{full}}(t)=4E_C\hat n^2+\frac{1}{2}E_L[\hat{\varphi}+\phi_{\mathrm{ext}}(t)]^2-E_J\cos\hat{\varphi},
\end{equation}
we write the controlled external flux as $\phi_{\mathrm{ext}}^{\mathrm{ctrl}}(t)=\pi+\phi_{\mathrm{off}}(t)$, where $\phi_{\mathrm{off}}(t)$ is the controlled flux offset from the half-flux sweet spot. In terms of the parametrization used in the main text, $\phi_{\mathrm{off}}(t)=\phi_{\mathrm{dc}}-\pi+\phi_{\mathrm{ac}}P(t)$. Equivalently, using the physical DC bias and normalized waveform, one can write $\phi_{\mathrm{off}}(t)=\tilde{\phi}_{\mathrm{dc}}-\pi+\tilde{\phi}_{\mathrm{ac}}\tilde P(t)$. The Hamiltonian can then be decomposed as
\begin{equation}
H_{\mathrm{full}}(t)=H_\pi+V_c(t),
\end{equation}
with $H_\pi=4E_C\hat n^2+\frac{1}{2}E_L(\hat{\varphi}+\pi)^2-E_J\cos\hat{\varphi}$, and $V_c(t)=E_L\phi_{\mathrm{off}}(t)\hat{\varphi}+\left[E_L\pi\phi_{\mathrm{off}}(t)+\frac{1}{2}E_L\phi_{\mathrm{off}}^2(t)\right]I$. The second term of $V_c(t)$ only produces an overall phase and does not affect the qubit dynamics. Hence, $V_c(t) \xrightarrow{\mathrm{drop}\ I} E_L\phi_{\mathrm{off}}(t)\hat{\varphi}$.
Throughout this subsection, the arrow $\xrightarrow{\mathrm{drop}\ I}$ denotes removing terms proportional to the identity operator in the qubit subspace.

Let $|\tilde g\rangle$ and $|\tilde e\rangle$ be the two lowest eigenstates of $H_\pi$, with eigenenergies $\tilde E_g$ and $\tilde E_e$, and define the two-level projector $\Pi_2=|\tilde g\rangle\langle\tilde g|+|\tilde e\rangle\langle\tilde e|$. In the static sweet-spot energy basis, we define $\tilde{\sigma}_z=|\tilde e\rangle\langle\tilde e|-|\tilde g\rangle\langle\tilde g|,\; 
\tilde{\sigma}_x=|\tilde g\rangle\langle\tilde e|+|\tilde e\rangle\langle\tilde g|,$
so that
\begin{equation}
\Pi_2H_\pi\Pi_2=\bar E I+\frac{\Delta}{2}\tilde{\sigma}_z
\xrightarrow{\mathrm{drop}\ I}
\frac{\Delta}{2}\tilde{\sigma}_z,\qquad
\Delta=\tilde E_e-\tilde E_g,\quad
\bar E=\frac{\tilde E_e+\tilde E_g}{2}.
\end{equation}

We next project the phase operator $\hat{\varphi}$. In the two-level subspace, the most general Hermitian projection of $\hat{\varphi}$ can be written as
\begin{equation}
\Pi_2\hat{\varphi}\Pi_2=C_\varphi I+D_\varphi\tilde{\sigma}_z+X_\varphi\tilde{\sigma}_x+Y_\varphi\tilde{\sigma}_y,
\end{equation}
where $\tilde{\sigma}_y=-i|\tilde g\rangle\langle\tilde e|+i|\tilde e\rangle\langle\tilde g|$. Equivalently, if $\varphi_{ij}=\langle\tilde i|\hat{\varphi}|\tilde j\rangle\; (i,j\in\{g,e\})$, then $C_\varphi=\frac{\varphi_{gg}+\varphi_{ee}}{2},\; D_\varphi=\frac{\varphi_{ee}-\varphi_{gg}}{2},\; \varphi_{ge}=X_\varphi-iY_\varphi.$
At the half-flux sweet spot, the nontrivial diagonal difference of the phase operator in the lowest two sweet-spot eigenstates vanishes by symmetry, so $D_\varphi=0$. The common diagonal component $C_\varphi I$ shifts both qubit levels equally and is therefore removed by $\xrightarrow{\mathrm{drop}\ I}$.

The remaining transverse part is determined by the off-diagonal matrix element $\varphi_{ge}$. Since the phases of the energy eigenstates are arbitrary, we may redefine $|\tilde g\rangle\rightarrow e^{i\alpha_g}|\tilde g\rangle$ and $|\tilde e\rangle\rightarrow e^{i\alpha_e}|\tilde e\rangle$. This transformation leaves $\Pi_2H_\pi\Pi_2$ diagonal, but changes the off-diagonal matrix element as $\varphi_{ge}\rightarrow e^{i(\alpha_e-\alpha_g)}\varphi_{ge}$. If $\varphi_{ge}=|\varphi_{ge}|e^{i\chi}$, choosing $\alpha_e-\alpha_g=\pi-\chi$ gives $\varphi_{ge}\rightarrow-|\varphi_{ge}|$. Therefore, in this phase convention, $Y_\varphi=-\mathrm{Im}\,\varphi_{ge}=0$ and $X_\varphi=\mathrm{Re}\,\varphi_{ge}=-\tilde{\varphi}_{ge}$, where $\tilde{\varphi}_{ge}=|\varphi_{ge}|=|\langle\tilde g|\hat{\varphi}|\tilde e\rangle|>0$. Thus,
\begin{equation}
\Pi_2\hat{\varphi}\Pi_2=C_\varphi I-\tilde{\varphi}_{ge}\tilde{\sigma}_x
\xrightarrow{\mathrm{drop}\ I}
-\tilde{\varphi}_{ge}\tilde{\sigma}_x .
\label{pro_phi}
\end{equation}
The $Y_\varphi\tilde{\sigma}_y$ term is not physically discarded; it is removed by choosing the relative phase of $|\tilde g\rangle$ and $|\tilde e\rangle$ such that the projected phase operator defines the $\tilde{\sigma}_x$ direction in the transverse plane.

Using Eq.~\eqref{pro_phi}, the nontrivial part of the projected controlled flux term $\Pi_2\left[E_L\phi_{\mathrm{off}}(t)\hat{\varphi}\right]\Pi_2$ becomes $-E_L\phi_{\mathrm{off}}(t)\tilde{\varphi}_{ge}\tilde{\sigma}_x$. 
Combining the nontrivial parts of $\Pi_2H_\pi\Pi_2$ and $\Pi_2V_c(t)\Pi_2$, the two-level Hamiltonian is
\begin{equation}
    H_q(t)=\frac{\Delta}{2}\tilde{\sigma}_z-E_L\tilde{\varphi}_{ge}\phi_{\mathrm{off}}(t)\tilde{\sigma}_x
    \label{H_q_2}
\end{equation}
Using $\phi_{\mathrm{off}}(t)=\tilde{\phi}_{\mathrm{dc}}-\pi+\tilde{\phi}_{\mathrm{ac}}\tilde P(t)$, one can define $\tilde B=2E_L(\tilde{\phi}_{\mathrm{dc}}-\pi)\tilde{\varphi}_{ge}$ and $\tilde A=E_L\tilde{\phi}_{\mathrm{ac}}\tilde{\varphi}_{ge}$, leading to the identical form $H_q(t)=\Delta\tilde{\sigma}_z/2-[\tilde B/2+\tilde A\tilde P(t)]\tilde{\sigma}_x$.

The main text uses the rotated Pauli convention $U_y=\exp\left(-i\frac{\pi}{4}\tilde{\sigma}_y\right)$, which leads to $U_y\tilde{\sigma}_zU_y^\dagger=\tilde{\sigma}_x,\;
U_y\tilde{\sigma}_xU_y^\dagger=-\tilde{\sigma}_z$.
After relabeling the rotated operators as $\tilde{\sigma}_z\rightarrow\sigma_x$ and $\tilde{\sigma}_x\rightarrow-\sigma_z$, the Hamiltonian becomes
\begin{equation}
H_q(t)=\frac{\Delta}{2}\sigma_x+\left[\frac{B}{2}+AP(t)\right]\sigma_z,
\end{equation}
which is Eq.~(3) of the main text. In this convention, $\sigma_x$ is the static sweet-spot splitting axis, while $\sigma_z$ is the two-level representation of the projected phase operator.

\subsection{System-bath coupling}

We now show that the same projected phase operator appears in the system-bath interaction. For $1/f$ flux noise, we write the external flux as $\phi_{\mathrm{ext}}(t)
\rightarrow \phi_{\mathrm{ext}}(t)+\delta\phi(t)$, where $\delta\phi(t)$ is a random flux fluctuation. Since only the inductive term in Eq.~(1) of the main text depends on $\phi_{\mathrm{ext}}(t)$, the first-order noise coupling is obtained by expanding
\begin{equation}
\begin{aligned}
\frac{1}{2}E_L[\hat{\varphi}+\phi_{\mathrm{ext}}(t)+\delta\phi(t)]^2=
\frac{1}{2}E_L[\hat{\varphi}+\phi_{\mathrm{ext}}(t)]^2+E_L[\hat{\varphi}+\phi_{\mathrm{ext}}(t)]\delta\phi(t)+O[\delta\phi^2(t)].
\end{aligned}
\end{equation}
The first term is the controlled Hamiltonian without random flux noise. In the second term, the part proportional to $\phi_{\mathrm{ext}}(t)\delta\phi(t)$ is an identity operator in the qubit Hilbert space and can be absorbed into the bath Hamiltonian or an overall phase. Thus, after absorbing $E_L\delta\phi(t)$ and coupling constants into the bath operator, the nontrivial system-bath coupling is proportional to $\hat{\varphi}\beta$.

Projecting this coupling to the qubit subspace and using Eq.~\eqref{pro_phi}, the nontrivial system operator is $-\tilde{\varphi}_{ge}\tilde{\sigma}_x$. Using the rotated Pauli convention $\tilde{\sigma}_x\rightarrow-\sigma_z$ and absorbing the remaining matrix element and sign into $\beta$, we obtain $H_{\mathrm{int}}=\sigma_z\beta$. Therefore, $\sigma_z$ in the system-bath interaction is the same two-level representation of the projected phase operator that appears in the driven qubit Hamiltonian.

For dielectric loss, we follow the effective noise model used in Refs.~\cite{Huang2021_SweetSpots,Cheng2022_Coherence}, in which this channel is also modeled as coupling through the same projected phase operator in the qubit subspace. Thus, flux noise and dielectric loss share the same system operator $\sigma_z$ in the effective two-level description, while their physical distinction is encoded in their different noise spectral densities.
}

\begin{figure}[t]
    \centering
    \includegraphics[width=.55\linewidth]{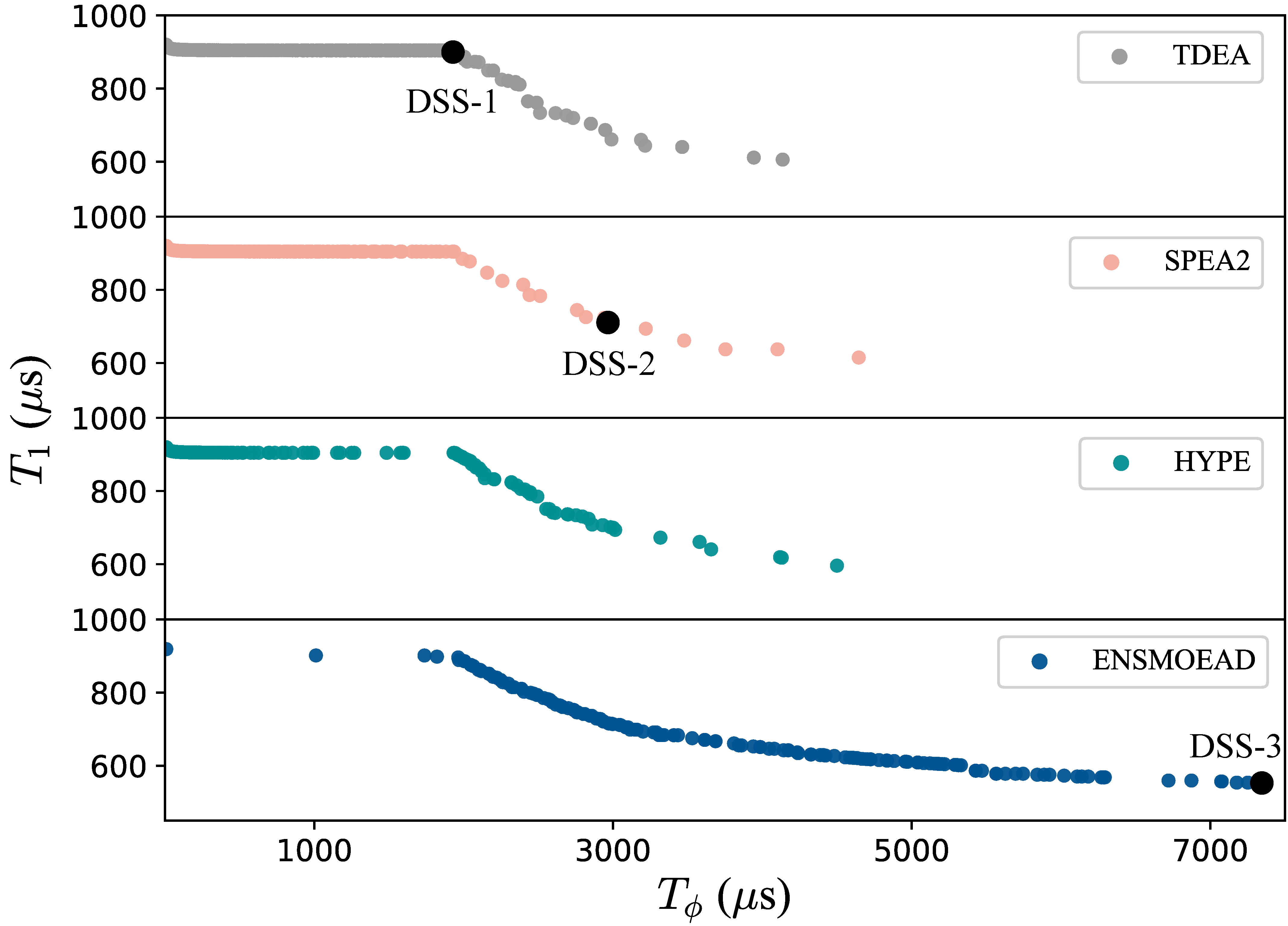}
    \caption{
        Pareto fronts obtained using four different multi-objective selection strategies (tDEA, SPEA2, ENS-MOEA/D, and HypE) under Fourier truncation length \(n=4\). Each PF represents the trade-off between \(T_1\) and \(T_\phi\) optimized via the periodic function \(P(t)\). The positions of three representative dynamical sweet spots (DSS 1–3), also listed in Table~\ref{Params_table}, are marked in each subfigure. All optimization runs were performed for 2000 iterations.
    }
    \label{PF_each}
\end{figure}
\begin{table*}[b]
    \centering
    \setlength{\tabcolsep}{4pt}
    \scriptsize
    \renewcommand{\arraystretch}{1.35}
    \setlength{\extrarowheight}{1.5pt}
    \resizebox{\textwidth}{!}{
    \begin{tabular}{c|ccccccccc}
        \hline\hline
        \textbf{Working point}
        & \textbf{\(p_0\)}
        & \textbf{\(p_1\)}
        & \textbf{\(p_2\)}
        & \textbf{\(p_3\)}
        & \textbf{\(p_4\)}
        & \textbf{\(\omega_d\)}
        & {\textbf{\(\tilde{\phi}_{\mathrm{dc}}\)}}
        & {\textbf{\(\tilde{\phi}_{\mathrm{ac}}\)}}
        & \textbf{Method}
        \\[4pt]
        \hline
        DSS-1
        & \(0.23\)
        & \(-0.55+0.21i\)
        & \(0.96-0.95i\)
        & \(-0.58+0.31i\)
        & \(0.14-0.85i\)
        & \(1.01\Omega_{ge}\)
        & {\(1.033\pi\)}
        & {\(0.094\pi\)}
        & tDEA
        \\
        DSS-2
        & \(0.69\)
        & \(0.73-0.99i\)
        & \(0.97-0.88i\)
        & \(0.28+0.84i\)
        & \(-0.16+0.58i\)
        & \(1.13\Omega_{ge}\)
        & {\(1.040\pi\)}
        & {\(0.096\pi\)}
        & SPEA2
        \\
        DSS-3
        & \(0.37\)
        & \(-0.99-1.00i\)
        & \(0.01-0.87i\)
        & \(-0.99+0.99i\)
        & \(0.99+1.00i\)
        & \(0.99\Omega_{ge}\)
        & {\(1.035\pi\)}
        & {\(0.109\pi\)}
        & ENS-MOEA/D
        \\
        Gate WP
        & \(0.11\)
        & \(-0.20+0.53i\)
        & \(-0.99-1.00i\)
        & \(0.43-0.18i\)
        & \(0.00+0.36i\)
        & \(0.98\Omega_{ge}\)
        & {\(1.032\pi\)}
        & {\(0.078\pi\)}
        & HypE
        \\
        \hline\hline
    \end{tabular}
    }
    \caption{
    {Modulation parameters for the three representative dynamical working points in Fig.~1 of the main text and for the gate working point (Gate WP). The gate working point is selected from the PF with truncation \(N=5\) and is used for the gate simulations.}
    }
    \label{Params_table}
\end{table*}

\section{Optimization Details of Pareto Fronts and Validation of Floquet Truncation}\label{sec:PF}

\subsection{Pareto Fronts Details}

The PFs obtained with four different selection strategies are shown in Fig.~\ref{PF_each}; their aggregated non-dominated set corresponds to Fig.~1(a) in the main text. 
{
The four strategies provide complementary coverage of the PF: SPEA2 and tDEA sample the high-\(T_1\), low-\(T_\phi\) side most densely, all four algorithms give broadly similar coverage in the intermediate part, and ENS-MOEA/D contributes predominantly to the large-\(T_\phi\), lower-\(T_1\) end. }
The locations of the three representative DSS working points listed in Table~I of the main text are marked in each subfigure, and their corresponding periodic flux-modulation parameters are given in Table~\ref{Params_table}.

\begin{figure*}[t]
    \centering
    \includegraphics[width=0.8\linewidth]{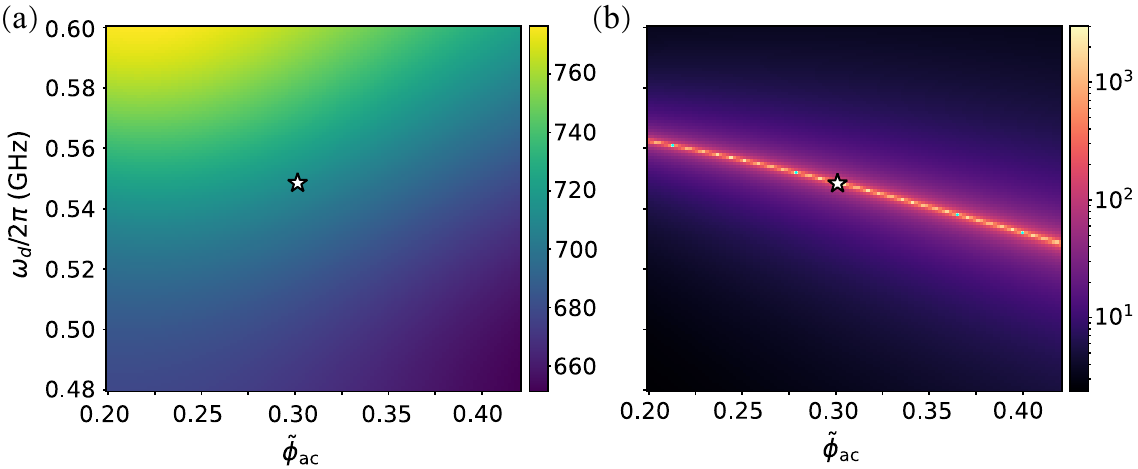}
    \caption{
    Two-dimensional physical-parameter scan around DSS-2.
    The optimized normalized waveform shape \(\tilde P(t)\) and the physical DC bias \(\tilde{\varphi}_{\rm dc}\) are fixed to the DSS-2 values, while the physical AC modulation amplitude \(\tilde{\varphi}_{\rm ac}\) and the modulation frequency \(\omega_d\) are varied over 
    \(\tilde{\varphi}_{\rm ac}\in[0.20,0.42]\) and 
    \(\omega_d/2\pi\in[0.48,0.60]~{\rm GHz}\).
    At each grid point, the coherence times are recomputed.
    (a) Energy relaxation time \(T_1\).
    (b) Pure-dephasing time \(T_\varphi\).
    The white star marks the DSS-2 operating point.
    }
    \label{fig:supp_DSS2_amp_freq_scan}
\end{figure*}

{
To further visualize the local coherence landscape in experimentally tunable parameters, we perform a two-dimensional scan around DSS-2 in the physical amplitude--frequency plane. 
In this scan, the optimized normalized waveform shape \(\tilde P(t)\) and the physical DC bias \(\tilde{\varphi}_{\rm dc}\) are fixed to the DSS-2 values, while the physical AC modulation amplitude \(\tilde{\varphi}_{\rm ac}\) and the modulation frequency \(\omega_d\) are varied. 
The scan window is 
\(\tilde{\varphi}_{\rm ac}\in[0.20,0.42]\) and 
\(\omega_d/2\pi\in[0.48,0.60]~{\rm GHz}\). 
For each grid point, we recompute the Floquet modes, quasienergy gap, filter coefficients, and the resulting coherence times \(T_1\) and \(T_\varphi\). 
The results are shown in Fig.~\ref{fig:supp_DSS2_amp_freq_scan}. 
The high-\(T_\varphi\) ridge in Fig.~\ref{fig:supp_DSS2_amp_freq_scan}(b) visualizes the local DSS structure around DSS-2 and shows that the enhanced dephasing time is not an isolated single-point feature.
}

\begin{figure}[t]
    \centering
    \includegraphics[width=.55\linewidth]{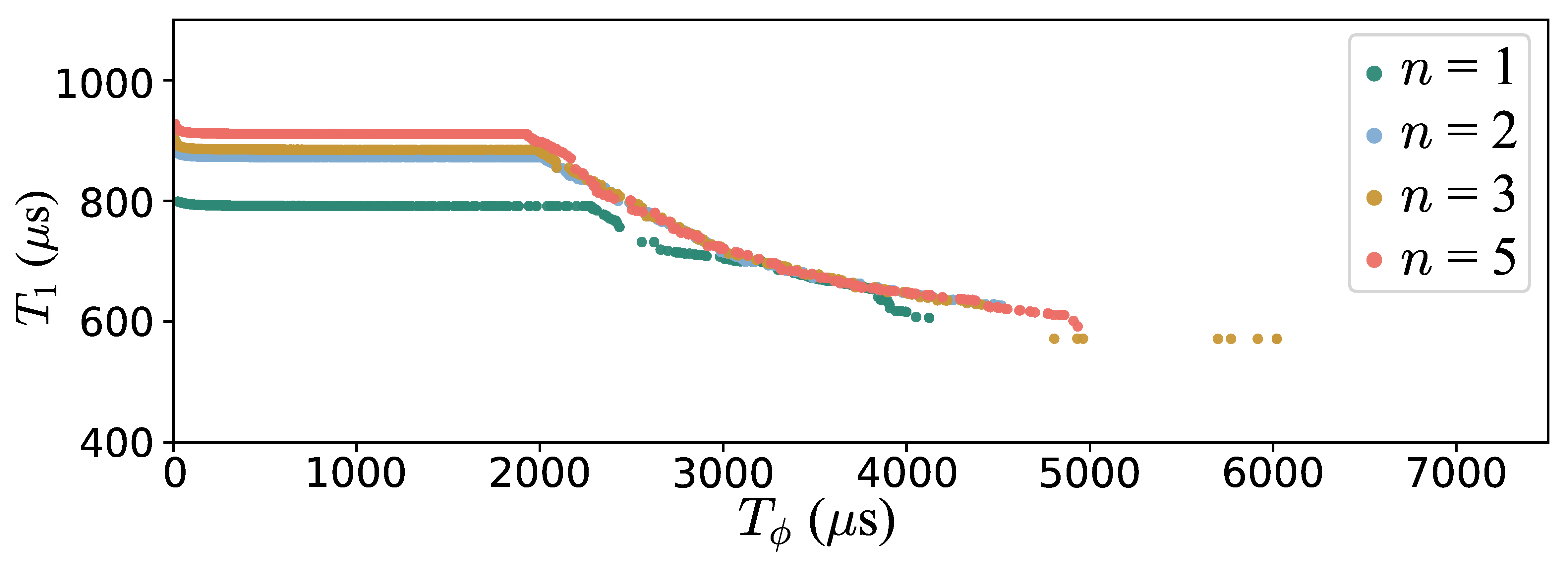}
    \caption{
        Comparison of Pareto fronts obtained under different Fourier truncation lengths \(n = 1\) to \(5\) for the periodic modulation function \(P(t)\). All four selection strategies were applied for each \(n\), with a maximum of 2000 optimization iterations. As \(n\) increases, the PFs become broader and more expressive, and higher values of \(T_1\) are observed. However, the case \(n = 5\) exhibits lower maximum \(T_\phi\) than the \(n = 4\) case (main text Fig.~1).
    }
    \label{PF_other}
\end{figure}

We employed the same four selection strategies to optimize \(T_1\) and \(T_\phi\) while varying the parameter length \(n\) of the periodic function \(P(t)=\sum_n p_n e^{in\omega_d t}\). Each strategy was run for up to 2000 iterations; the resulting PFs are shown in Fig.~\ref{PF_other}. 
For all tested values of \(n\), the PFs lie in the ranges \(T_\phi \in [2000,7500]\) and \(T_1 \in [500,900]\). The attainable maximum \(T_1\) generally increases with \(n\). Notably, the maximal \(T_\phi\) observed at \(n=5\) is lower than that reported for \(n=4\) in Fig.~1 of the main text. We attribute this discrepancy to incomplete convergence at \(n=5\): enlarging \(n\) increases the dimensionality and complexity of the parameter space, which typically requires more iterations to reach comparable convergence. Due to computational-resource limits we did not extend the optimizations for larger \(n\). It is expected that further improvements in the maximum achievable \(T_\phi\) may be obtained with larger $n$.

\subsection{Numerical Errors Induced by Floquet Matrix Truncation}\label{sec:Floquet_Matrix}

In this section we quantify the numerical errors introduced by truncating the Floquet matrix \(\mathcal{H}_F\). 
As shown in Sec.~II.B of the main text, the decoherence rates \(\gamma_{\pm,z}\) depend on the Floquet states \(|\omega_{\pm}(t)\rangle\) and their quasienergies \(\epsilon_{\pm}\); consequently, the accuracy of the computed decoherence rates is governed by the numerical precision of these Floquet quantities. We adopt a truncation-free reference Floquet basis \(|\omega^{U}_\pm(t)\rangle\), obtained by diagonalizing the propagator \(U_q(t)\) (i.e. the time-ordered evolution operator over one drive period)~\cite{Creffield2003_Floquet}. This avoids constructing or truncating \(\mathcal{H}_F\) and is limited only by the numerical accuracy of the differential-equation solver, so we use it to quantify truncation-induced errors.

\begin{figure}[h]
    \centering
    \includegraphics[width=0.55\linewidth]{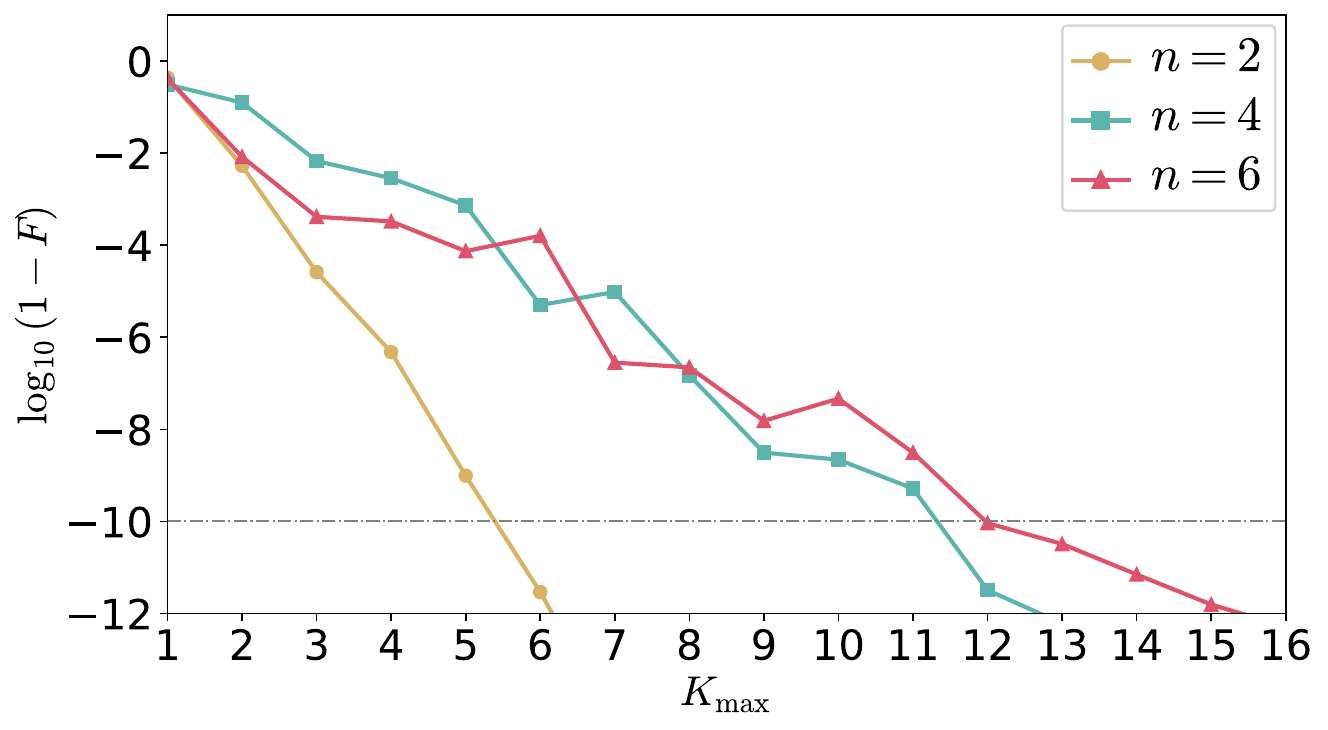}
    \caption{
        Numerical fidelity between the Floquet state \(|\omega^{H}_+(0)\rangle\) and the reference Floquet state \(|\omega^{U}_+(0)\rangle\). The infidelity quantifies the truncation-induced error for different truncation dimensions \(K_{\max}\). The numerical error decreases exponentially with increasing \(K_{\max}\), and falls below \(10^{-10}\) when \(K_{\max} \geq 3n\), justifying the truncation setting used in the main text.
    }
    \label{floquet_dim}
\end{figure}

We quantify truncation error using the fidelity $F = \left|\langle \omega^{U}_+(0) \mid \omega^{H}_+(0)\rangle\right|$, where \(|\omega^{H}_+(0)\rangle\) is obtained by diagonalizing the truncated Floquet matrix \(\mathcal{H}_F\) with truncation dimension \(2K_{\max}+1\), and \(|\omega^{U}_+(0)\rangle\) is the truncation-free reference state. When the parameter length \(n\) of the periodic function \(P(t)\) increases, the Floquet matrix acquires more nonzero blocks, which can lead to larger truncation-induced errors. We therefore compute the fidelity under different truncation settings to characterize this effect. The numerical results are plotted in Fig.~\ref{floquet_dim}. The infidelity \(1-F\) drops below \(10^{-10}\) once the truncation dimension satisfies \(K_{\max}=3n\).

{
\section{Multilevel full-fluxonium validation}
\label{sec:multilevel_validation}

\subsection{\(N\)-level decoherence rates and convergence of the DSS condition}
\label{sec:multilevel_rates_convergence}

We use the same notation and noise conventions as in Sec.~\ref{sec:two_level_projection}, and only give the corresponding \(N\)-level generalization. Let
\(\{|\tilde{0}\rangle,|\tilde{1}\rangle,\ldots,|\widetilde{N-1}\rangle\}\)
denote the lowest \(N\) static eigenstates of \(H_\pi\), namely the full fluxonium Hamiltonian at the half-flux sweet spot \(\phi_{\rm ext}=\pi\). We define the \(N\)-level projector
\(\Pi_N=\sum_{j=0}^{N-1}|\tilde j\rangle\langle\tilde j| .\)
In this truncated basis, the static Hamiltonian and the projected phase operator are
\begin{equation}
    D_N
    =
    \Pi_NH_\pi\Pi_N
    =
    \mathrm{diag}(E_0,E_1,\ldots,E_{N-1}),
    \qquad
    \hat{\varphi}_N
    =
    \Pi_N\hat{\varphi}\Pi_N
    =
    \sum_{i,j=0}^{N-1}
    \varphi_{ij}
    |\tilde{i}\rangle\langle\tilde{j}| ,
    \label{eq:phiN_def}
\end{equation}
where \(\varphi_{ij}=\langle\tilde{i}|\hat{\varphi}|\tilde{j}\rangle\). Following the two-level projection in Sec.~\ref{sec:two_level_projection}, the controlled flux offset from the half-flux sweet spot is denoted by \(\phi_{\mathrm{off}}(t)\), so that
\(\phi_{\mathrm{ext}}^{\mathrm{ctrl}}(t)=\pi+\phi_{\mathrm{off}}(t)\). After dropping identity terms, the \(N\)-level controlled Hamiltonian is
\begin{equation}
    H_N(t)
    =
    D_N
    +
    E_L\phi_{\mathrm{off}}(t)\hat{\varphi}_N .
    \label{eq:HN_multilevel}
\end{equation}
This is the direct multilevel counterpart of Eq.~\eqref{H_q_2}.  Similarly, the corresponding \(N\)-level interaction Hamiltonian of the system-bath coupling is
\begin{equation}
    H_{\mathrm{int}}^{(N)}(t)
    =
    \hat{\varphi}_N\beta(t),
    \label{eq:Hint_multilevel}
\end{equation}
where \(\beta(t)\) is the effective bath variable. The constant coupling factors, including \(E_L\), are absorbed into the noise spectrum \(S(\omega)\) used below, as in the two-level derivation.

The \(N\)-level Floquet states are written as $|\psi_{\pm}^{(N)}(t)\rangle = e^{-i\epsilon_{\pm}^{(N)}t}    |\omega_{\pm}^{(N)}(t)\rangle ,$
where \(|\omega_{\pm}^{(N)}(t)\rangle\) are periodic Floquet modes. In the multilevel Floquet spectrum, we choose the two Floquet branches corresponding to the driven continuation of the two lowest static fluxonium states. Their quasienergy gap is \(\Omega_N=\epsilon_+^{(N)}-\epsilon_-^{(N)} .\)
Applying the same Born-frequency decomposition as in the two-level Floquet derivation to Eq.~\eqref{eq:Hint_multilevel}, the only required replacement is
\(\sigma_z\rightarrow\hat{\varphi}_N\). We define the Fourier components of the multilevel noise-coupling matrix elements as
\begin{equation}
    V_{\alpha\beta,N}^{[k]}
    =
    \frac{1}{T}
    \int_0^T dt\,
    e^{ik\omega_d t}
    \langle
    \omega_{\alpha}^{(N)}(t)
    |
    \hat{\varphi}_N
    |
    \omega_{\beta}^{(N)}(t)
    \rangle ,
    \qquad
    \alpha,\beta\in\{+,-\}.
    \label{eq:Vab_multilevel}
\end{equation}
The corresponding \(N\)-level filter coefficients are $g_{z,N}^{[k]} =\left({ V_{++,N}^{[k]}-V_{--,N}^{[k]}}\right)/2,\;
    g_{+,N}^{[k]}=V_{+-,N}^{[k]},$ and $
    g_{-,N}^{[k]}=V_{-+,N}^{[k]}$.
The effective relaxation rates in the logical Floquet subspace are then
\begin{equation}
    \gamma_{+,N}    =    \sum_k
    |g_{+,N}^{[k]}|^2
    S(k\omega_d-\Omega_N),
    \qquad
    \gamma_{-,N}=\sum_k
    |g_{-,N}^{[k]}|^2
    S(k\omega_d+\Omega_N),
    \label{eq:gamma_pm_multilevel}
\end{equation}
and $T_1^{(N)} = 1/{(\gamma_{+,N}+\gamma_{-,N})}$.
The pure-dephasing rate is
\begin{equation}
    \gamma_{\phi,N}
    =
    A_f |g_{z,N}^{[0]}|
    \sqrt{2|\ln(\omega_{\mathrm{ir}}t_m)|}
    +
    \sum_{k\neq0}
    |g_{z,N}^{[k]}|^2
    S(k\omega_d),
    \label{eq:gamma_phi_multilevel}
\end{equation}
and $T_\phi^{(N)}=1/{\gamma_{\phi,N}}$.
Equations~\eqref{eq:gamma_pm_multilevel}--\eqref{eq:gamma_phi_multilevel} reduce to the two-level expressions when \(N=2\).

\begin{figure}[tb]
    \centering
    \includegraphics[width=\linewidth]{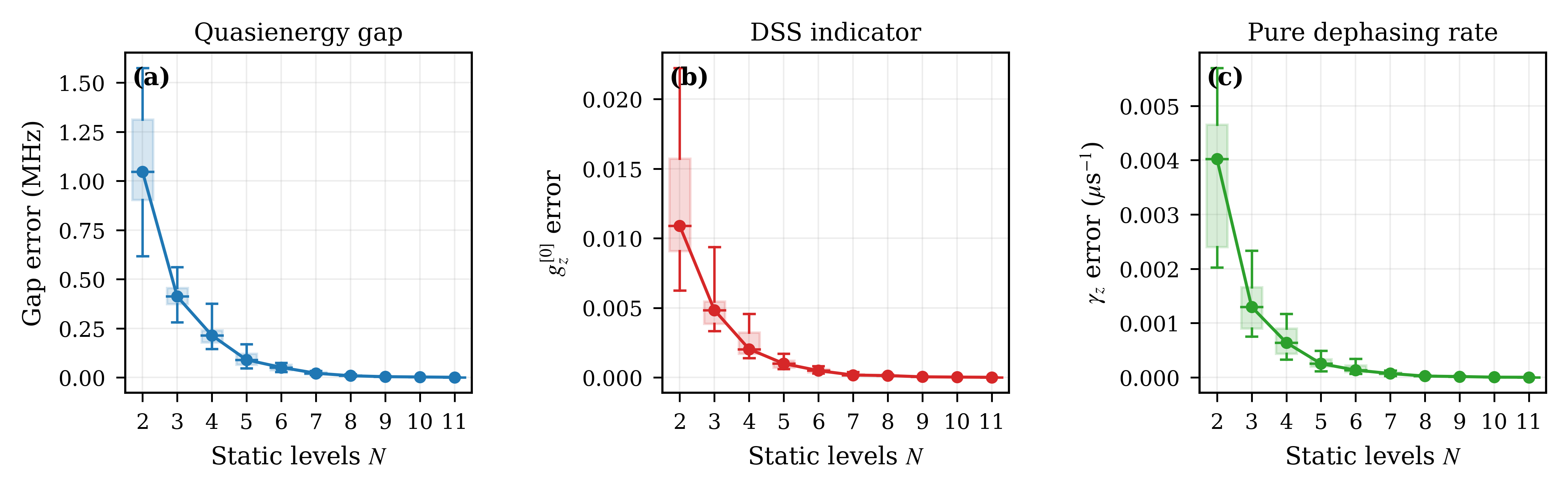}
    \caption{
    Convergence of multilevel Floquet quantities with respect to the static-level truncation \(N\).
    We generate 100 random flux waveforms and compute, for each waveform and each \(N\), the error relative to the corresponding \(N=50\) result.
    (a) Error of the quasienergy gap.
    (b) Error of the DSS indicator \(|g_{z,N}^{[0]}|\).
    (c) Error of the pure-dephasing rate \(\gamma_{\phi,N}\).
    The boxes show the distribution over random waveforms, and the solid lines connect the median values.
    The three quantities become stable for \(N\ge 5\), indicating that the DSS condition is converged at the \(N=5\) full-fluxonium truncation.
    }
    \label{fig:multilevel_gap_gz_gamma_convergence}
\end{figure}

The pure-dephasing time is especially sensitive to \(g_{z,N}^{[0]}\). From Eq.~\eqref{eq:HN_multilevel}, the Floquet Hellmann--Feynman relation gives
\(\frac{\partial\Omega_N}{\partial\tilde{\phi}_{\mathrm{dc}}}
=E_L\left(V_{++,N}^{[0]}-V_{--,N}^{[0]}\right)
=2E_Lg_{z,N}^{[0]}\).
Thus, the DSS condition in the \(N\)-level model can be equivalently monitored by the zero-frequency diagonal filter coefficient, $\frac{\partial\Omega_N}    {\partial\tilde{\phi}_{\mathrm{dc}}}  \approx0,$ or equivalently $|g_{z,N}^{[0]}|\approx0$.
A small truncation-induced shift of \(\Omega_N\) can therefore move the zero of
\(\partial\Omega_N/\partial\tilde{\phi}_{\mathrm{dc}}\), and hence shift the DSS location in parameter space.

To quantify the convergence with respect to the level truncation, we randomly generate 100 flux waveforms within the same parameter domain used in the PF optimization. For each waveform, we compute the quasienergy gap \(\Omega_N\), the DSS indicator \(|g_{z,N}^{[0]}|\), and the pure-dephasing rate \(\gamma_{\phi,N}\) at different truncation dimensions \(N\). The corresponding \(N=50\) results are used as the reference values, and the absolute errors are evaluated for each random waveform. Figure~\ref{fig:multilevel_gap_gz_gamma_convergence} shows the resulting error distributions as box plots, with the median values connected by solid lines. The median errors decrease rapidly with \(N\), and all three quantities become stable for \(N\ge 5\). This indicates that the quasienergy gap, the DSS condition, and the pure-dephasing rate are sufficiently converged at the \(N=5\) full-fluxonium truncation for the parameters considered here.

\subsection{\(N=5\) PF and higher-\(N\) validation}
\label{sec:N5_PF_validation}

Guided by the convergence analysis above, we repeat the PF optimization using the five-level truncation of the fluxonium Hamiltonian. The optimization protocol is kept the same as in the main text: ENS-MOEA/D, HypE, SPEA2, and tDEA are run independently, and the final populations from all algorithms and all runs are merged and subjected to a final non-dominated sorting to obtain the aggregated PF.

The resulting \(N=5\) PF is shown in Fig.~\ref{fig:N5_truncated_PF}. It remains nearly unchanged in the objective space compared with the two-level PF shown in the main text. This indicates that higher-level corrections mainly shift the optimized waveform parameters and the DSS locations in parameter space, while leaving the achievable \(T_1\)--\(T_\phi\) trade-off qualitatively unchanged.

\begin{figure}[t]
    \centering
    \includegraphics[width=0.55\linewidth]{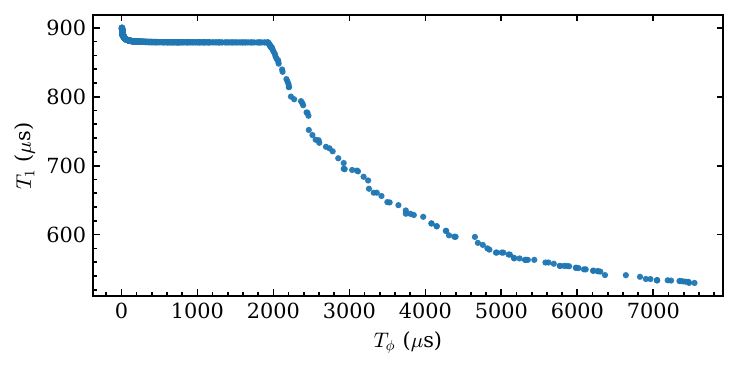}
    \caption{
    Aggregated Pareto front obtained by repeating the optimization with the five-level truncation of the fluxonium Hamiltonian. The PF remains nearly unchanged in the objective space compared with the two-level PF in the main text, showing that higher-level corrections do not qualitatively alter the achievable \(T_1\)--\(T_\phi\) trade-off.
    }
    \label{fig:N5_truncated_PF}
\end{figure}

\begin{figure}[t]
    \centering
    \includegraphics[width=\linewidth]{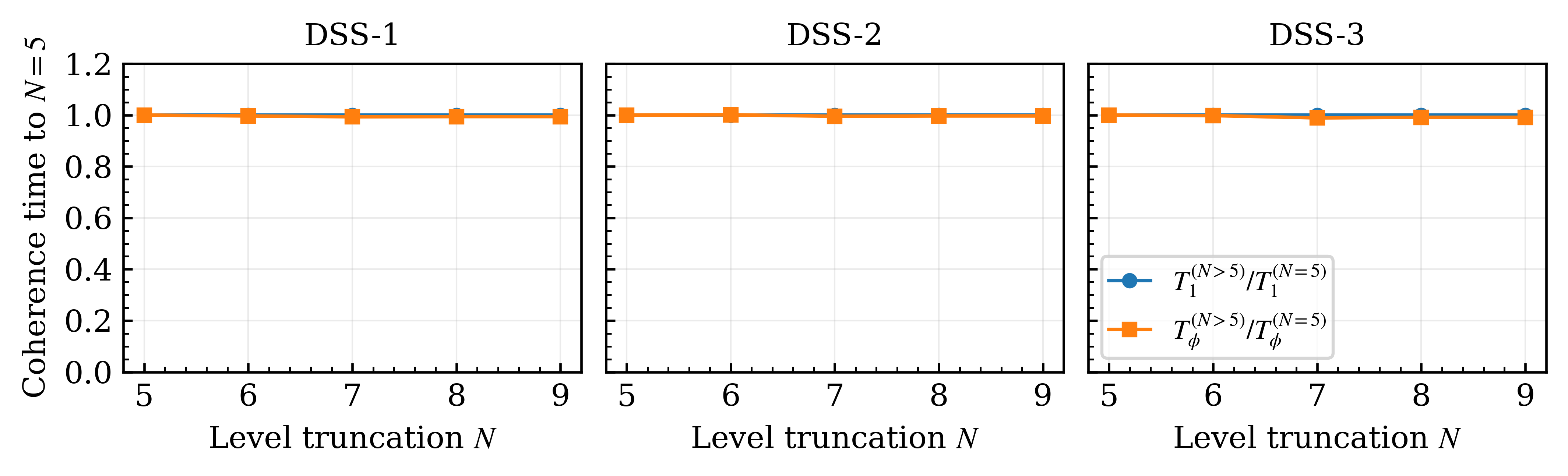}
    \caption{
    Higher-\(N\) validation of three representative DSS points selected from the five-level PF. The plotted quantities are \(T_1(N)/T_1(N=5)\) and \(T_\phi(N)/T_\phi(N=5)\). Both coherence times remain stable under further increase of \(N\), confirming that the \(N=5\) truncation is sufficient for the reported multilevel validation.
    }
    \label{fig:N5_DSS_higher_N_validation}
\end{figure}

We select three representative DSS points from the \(N=5\) PF. 
Their coherence times, \(T_1\) and \(T_\phi\), are consistent with those of the three representative DSS points reported in Table~I of the main text. Then, the physical waveforms are fixed, and \(T_1^{(N)}\) and \(T_\phi^{(N)}\) are recomputed using the \(N\)-level decoherence-rate expressions in Sec.~\ref{sec:multilevel_rates_convergence}. To make the convergence clear despite the different absolute scales of \(T_1\) and \(T_\phi\), we plot the normalized ratios $\frac{T_1(N\geq5)}{T_1(N=5)},\; \frac{T_\phi(N\geq5)}{T_\phi(N=5)}$. The results are shown in Fig.~\ref{fig:N5_DSS_higher_N_validation}. For all three selected working points, both ratios remain close to unity when the truncation dimension is increased beyond \(N=5\). This confirms that the \(N=5\) PF points are stable under further increase of the static-level truncation.

In summary, higher fluxonium levels introduce quantitative corrections to the decoherence rates and shift the numerical DSS locations in parameter space. However, after re-optimizing the PF with the converged five-level truncation, the PF in the objective space remains essentially unchanged, and representative DSS points remain stable under higher-\(N\) validation. Therefore, these multilevel corrections do not affect the main qualitative conclusions of the Pareto-front engineering framework.

}

\section{Upper bound of $T_1$ and Applicability to modified noise spectra}

\subsection{Upper bound of $T_1$}\label{sec:proof_Theorem}

We prove the global upper bound on \(T_1\) stated in Theorem~1. The proof consists of two steps. First, we show that the symmetrized spectrum \(\tilde S(\omega)=S(\omega)+S(-\omega)\) has a positive global lower bound under the noise model used in Eq.~(12) of the main text. Second, we show that the DSS condition gives a positive lower bound on the total transition-filter weight \(\sum_k|g_+^{[k]}|^2\).

We start from the long-time Floquet--Markov rates in Eq.~(13) of the main text. The total relaxation rate can be written as
\begin{align}\label{target1}
\gamma_1
&= \sum_{k\in\mathbb{Z}} |g_{+}^{[k]}|^2 S(k\omega_d-\Omega)
  + \sum_{k\in\mathbb{Z}}  |g_{-}^{[k]}|^2 S(k\omega_d+\Omega) \nonumber\\
&= \sum_{k\in\mathbb{Z}} |g_{+}^{[k]}|^2\big(S(k\omega_d-\Omega) + S(-k\omega_d+\Omega)\big) \nonumber\\
&\equiv \sum_{k\in\mathbb{Z}} |g_{+}^{[k]}|^2 \,\tilde{S}(k\omega_d-\Omega),
\end{align}
where in the second line we used \(g_-(t)=g_+^*(t)\), which gives \(|g_-^{[k]}|=|g_+^{[-k]}|\), and then relabeled the summation index. Thus, a lower bound on \(\gamma_1\) follows from a lower bound on \(\tilde S(\omega)\) and a lower bound on \(\sum_k|g_+^{[k]}|^2\).

We first bound the symmetrized spectrum. From Eq.~(12) of the main text, \(S(\omega)=A_f^2|2\pi/\omega|+\kappa(\omega,\mathcal T)A_d(\hbar\omega/2\pi)^2\). Therefore,
\begin{equation}
\begin{split}
\tilde S(\omega)
&=S(\omega)+S(-\omega)\\
&=A_f^2\left|\frac{2 \pi}{\omega}\right|+A_f^2\left|\frac{2 \pi}{-\omega}\right|
+\kappa(\omega, \mathcal T) A_d\left(\frac{\hbar \omega}{2 \pi}\right)^2
+\kappa(-\omega, \mathcal T) A_d\left(\frac{-\hbar \omega}{2 \pi}\right)^2\\
&=\frac{4\pi A_f^2}{|\omega|}
+
\left[\kappa(\omega,\mathcal T)+\kappa(-\omega,\mathcal T)\right]
A_d\left(\frac{\hbar\omega}{2\pi}\right)^2 .
\end{split}
\end{equation}
The thermal factor satisfies
\begin{equation}
\begin{split}
\kappa(\omega,\mathcal T)+\kappa(-\omega,\mathcal T)
&=\frac{1}{2}\left|1+\coth\left(\frac{\hbar\omega}{2k_B\mathcal T}\right)\right|
+\frac{1}{2}\left|1-\coth\left(\frac{\hbar\omega}{2k_B\mathcal T}\right)\right|\\
&=\left|\coth\left(\frac{\hbar\omega}{2k_B\mathcal T}\right)\right|\ge 1 .
\end{split}
\end{equation}
Hence, \(\tilde S(\omega)\ge 4\pi A_f^2/|\omega|+A_d(\hbar\omega/2\pi)^2\). 
{
To obtain a global constant lower bound, we split the first term as \(4\pi A_f^2/|\omega|=2\pi A_f^2/|\omega|+2\pi A_f^2/|\omega|\). Applying the three-term arithmetic--geometric mean inequality, \(x_1+x_2+x_3\ge 3(x_1x_2x_3)^{1/3}\), gives
\begin{equation}
\begin{split}
\tilde S(\omega)
\ge
\frac{2\pi A_f^2}{|\omega|}
+
\frac{2\pi A_f^2}{|\omega|}
+
A_d\left(\frac{\hbar\omega}{2\pi}\right)^2
\ge
3\left[
\frac{2\pi A_f^2}{|\omega|}
\cdot
\frac{2\pi A_f^2}{|\omega|}
\cdot
A_d\left(\frac{\hbar\omega}{2\pi}\right)^2
\right]^{1/3}
=
3(A_f^4A_d\hbar^2)^{1/3}.
\label{S_lower_global}
\end{split}
\end{equation}
This global lower bound is independent of the sampled sideband frequency.

Next, we prove the lower bound on \(\sum_k|g_+^{[k]}|^2\) under the DSS condition. We use the interaction-picture operator defined in the main text, \(\sigma_z(t)=U_q^\dagger(t)\sigma_zU_q(t)\). In the floquet basis $|\omega_\pm\rangle$, its Bohr-frequency decomposition is
\begin{equation}
\begin{split}
\sigma_z(t)
&=
\frac{1}{2}\tau_z(0)\operatorname{Tr}[\sigma_z\tau_z(t)]
+
\tau_+(0)e^{+i\Omega t}\operatorname{Tr}[\sigma_z\tau_-(t)]
+
\tau_-(0)e^{-i\Omega t}\operatorname{Tr}[\sigma_z\tau_+(t)]\\
&\equiv
g_z(t)\tau_z(0)
+
g_+(t)e^{i\Omega t}\tau_+(0)
+
g_+^*(t)e^{-i\Omega t}\tau_-(0)\\
&=
\left(
\begin{array}{cc}
g_z(t) & g_+(t)e^{i\Omega t}\\
g_+^*(t)e^{-i\Omega t} & -g_z(t)
\end{array}
\right).
\end{split}
\end{equation}
Here \(g_z(t)=\frac{1}{2}\operatorname{Tr}[\sigma_z\tau_z(t)]\), \(g_+(t)=\operatorname{Tr}[\sigma_z\tau_-(t)]\), and \(g_-(t)=\operatorname{Tr}[\sigma_z\tau_+(t)]=g_+^*(t)\) are periodic functions. 
Since \(\sigma_z^2(t)=I\), the diagonal elements of the above matrix representation satisfy \(g_z^2(t)+|g_+(t)|^2=1\). Therefore, \(|g_+(t)|^2=1-g_z^2(t)\).
We now write \(\tau_z(t)\) in Bloch form as \(\tau_z(t)=\mathbf n(t)\cdot\bm\sigma\), where \(|\mathbf n(t)|=1\). Since \(g_z(t)=\frac{1}{2}\operatorname{Tr}[\sigma_z\tau_z(t)]\), one obtains
\begin{equation}
\begin{split}
g_z(t)
=
\frac{1}{2}\operatorname{Tr}[\sigma_z(n_x\sigma_x+n_y\sigma_y+n_z\sigma_z)]
=
n_z(t).
\end{split}
\end{equation}
Therefore, \(|g_+(t)|^2=1-n_z^2(t)\). By Parseval's identity, for any periodic function \(f(t)\) with Fourier coefficients \(f^{[k]}=\frac{1}{T_d}\int_0^{T_d}f(t)e^{ik\omega_dt}dt\), one has \(\frac{1}{T_d}\int_0^{T_d}|f(t)|^2dt=\sum_k|f^{[k]}|^2\). Applying this identity to \(g_+(t)\) gives
\begin{equation}
\begin{split}
\sum_k |g_+^{[k]}|^2
=
\frac{1}{T_d}\int_0^{T_d}|g_+(t)|^2dt
=
1-\frac{1}{T_d}\int_0^{T_d}n_z^2(t)dt
\equiv
1-\langle n_z^2(t)\rangle ,
\label{parseval_gplus}
\end{split}
\end{equation}
where \(\langle\cdots\rangle\) denotes the average over one driving period.

The ideal DSS condition is \(g_z^{[0]}=0\). Since \(g_z(t)=n_z(t)\), this condition is equivalent to \(g_z^{[0]}=\frac{1}{T_d}\int_0^{T_d}g_z(t)dt=\frac{1}{T_d}\int_0^{T_d}n_z(t)dt=\langle n_z(t)\rangle=0\). Thus, the Fourier series of \(n_z(t)\) has no \(k=0\) component, namely \(n_z(t)=\sum_{k\neq0}n_z^{[k]}e^{-ik\omega_dt}\). Applying Parseval's identity to \(n_z(t)\) and \(\dot n_z(t)\), we obtain \(\langle n_z^2(t)\rangle=\sum_{k\neq0}|n_z^{[k]}|^2\) and \(\langle\dot n_z^2(t)\rangle=\sum_{k\neq0}k^2\omega_d^2|n_z^{[k]}|^2\ge\omega_d^2\sum_{k\neq0}|n_z^{[k]}|^2\). Therefore,
\begin{equation}
\langle n_z^2(t)\rangle
\le
\frac{1}{\omega_d^2}\langle\dot n_z^2(t)\rangle .
\label{wirtinger}
\end{equation}

It remains to bound \(\dot n_z(t)\). Starting from the Schrödinger equation \(i\frac{d}{dt}|\psi_\pm(t)\rangle=H_q(t)|\psi_\pm(t)\rangle\) and using the Floquet form \(|\psi_\pm(t)\rangle=e^{-i\epsilon_\pm t}|\omega_\pm(t)\rangle\), we get \(i\frac{d}{dt}|\omega_\pm(t)\rangle=[H_q(t)-\epsilon_\pm]|\omega_\pm(t)\rangle\). For the projectors \(P_\pm(t)=|\omega_\pm(t)\rangle\langle\omega_\pm(t)|\), this gives
\begin{equation}
\begin{split}
\dot P_\pm(t)
&=
\frac{d}{dt}|\omega_\pm(t)\rangle\langle\omega_\pm(t)|
+
|\omega_\pm(t)\rangle\frac{d}{dt}\langle\omega_\pm(t)|\\
&=
-i[H_q(t)-\epsilon_\pm]P_\pm(t)
+
iP_\pm(t)[H_q(t)-\epsilon_\pm]\\
&=
-i[H_q(t),P_\pm(t)].
\end{split}
\end{equation}
Since \(\tau_z(t)=P_+(t)-P_-(t)\), one has \(\dot\tau_z(t)=-i[H_q(t),\tau_z(t)]\). The effective two-level Hamiltonian is \(H_q(t)=\frac{\Delta}{2}\sigma_x+\lambda(t)\sigma_z\), where \(\lambda(t)=\frac{B}{2}+AP(t)\). Equivalently, \(H_q(t)=\frac{1}{2}\mathbf h(t)\cdot\bm\sigma\) with \(\mathbf h(t)=(\Delta,0,2\lambda(t))\). Substituting \(\tau_z(t)=\mathbf n(t)\cdot\bm\sigma\) into \(\dot\tau_z(t)=-i[H_q(t),\tau_z(t)]\), and using \([\mathbf a\cdot\bm\sigma,\mathbf b\cdot\bm\sigma]=2i(\mathbf a\times\mathbf b)\cdot\bm\sigma\), gives
\begin{equation}
\begin{split}
\dot{\mathbf n}(t)\cdot\bm\sigma
=
-i\left[\frac{1}{2}\mathbf h(t)\cdot\bm\sigma,\mathbf n(t)\cdot\bm\sigma\right]
=
-i\left[i(\mathbf h(t)\times\mathbf n(t))\cdot\bm\sigma\right]
=
(\mathbf h(t)\times\mathbf n(t))\cdot\bm\sigma .
\end{split}
\end{equation}
Thus \(\dot{\mathbf n}(t)=\mathbf h(t)\times\mathbf n(t)\). Taking the \(z\) component gives \(\dot n_z=h_xn_y-h_yn_x=\Delta n_y\), because \(h_x=\Delta\) and \(h_y=0\). Hence, \(\dot n_z^2=\Delta^2n_y^2\). Since \(|\mathbf n(t)|^2=1\), one has \(n_y^2\le1-n_z^2\), and therefore
\begin{equation}
\langle\dot n_z^2(t)\rangle
\le
\Delta^2\left[1-\langle n_z^2(t)\rangle\right].
\label{ndot_bound}
\end{equation}

Combining Eq.~\eqref{wirtinger} and Eq.~\eqref{ndot_bound}, we find \(\langle n_z^2(t)\rangle\le\frac{\Delta^2}{\omega_d^2}\left[1-\langle n_z^2(t)\rangle\right]\), which implies \(\langle n_z^2(t)\rangle\le\frac{\Delta^2}{\omega_d^2+\Delta^2}\). Using Eq.~\eqref{parseval_gplus}, this gives
\begin{equation}
\sum_k|g_+^{[k]}|^2
\ge
\frac{\omega_d^2}{\omega_d^2+\Delta^2}.
\label{gplus_lower}
\end{equation}
Finally, substituting Eq.~\eqref{S_lower_global} and Eq.~\eqref{gplus_lower} into Eq.~\eqref{target1}, we obtain \(\gamma_1\ge 3(A_f^4A_d\hbar^2)^{1/3}\frac{\omega_d^2}{\omega_d^2+\Delta^2}\). Therefore,
\begin{equation}
T_1=\frac{1}{\gamma_1}
\le
\frac{\omega_d^2+\Delta^2}
{3\omega_d^2(A_f^4A_d\hbar^2)^{1/3}}.
\end{equation}

\noindent
{\bf Remark}\; (Away from the DSS).
The DSS condition is the step that turns the lower bound on the transition-filter weight into a waveform-independent constant. Without imposing \(g_z^{[0]}=0\), one can repeat the same Fourier argument for \(n_z(t)-g_z^{[0]}\), which has zero average because \(g_z^{[0]}=\langle n_z(t)\rangle\). Following the derivation leading to Eq.~\eqref{wirtinger}, we obtain
\(\left\langle \left[n_z(t)-g_z^{[0]}\right]^2\right\rangle
\le
\frac{1}{\omega_d^2}\left\langle \dot n_z^2(t)\right\rangle .\)
Since \(g_z^{[0]}\) is the average of \(n_z(t)\), the left-hand side is \(\langle n_z^2(t)\rangle-|g_z^{[0]}|^2\). Combining this result with Eq.~\eqref{ndot_bound} gives
\(\langle n_z^2(t)\rangle-|g_z^{[0]}|^2
\le
\frac{\Delta^2}{\omega_d^2}
\left[1-\langle n_z^2(t)\rangle\right].\)
Rearranging the above inequality yields
\(\langle n_z^2(t)\rangle
\le
\frac{\Delta^2+\omega_d^2|g_z^{[0]}|^2}
{\omega_d^2+\Delta^2}.\)
Using Eq.~\eqref{parseval_gplus}, we therefore find
\(\sum_k |g_+^{[k]}|^2
\ge
\frac{\omega_d^2}{\omega_d^2+\Delta^2}
\left(1-|g_z^{[0]}|^2\right).\)
Together with the spectral lower bound in Eq.~\eqref{S_lower_global}, this gives the formal upper bound
\begin{equation}
    T_1
\le
\frac{\omega_d^2+\Delta^2}
{3\omega_d^2(A_f^4A_d\hbar^2)^{1/3}
\left(1-|g_z^{[0]}|^2\right)}
\end{equation}
whenever \(|g_z^{[0]}|<1\). This expression reduces to Theorem~1 in the ideal DSS limit \(g_z^{[0]}=0\). Away from the DSS, however, \(|g_z^{[0]}|\) can approach unity, in which case the bound becomes arbitrarily loose. Thus, a nontrivial waveform-independent upper bound is guaranteed only in the DSS regime.

\subsection{Applicability to modified noise spectra}\label{sec:modified_noise}

We briefly discuss how the proof extends to modified noise spectra, assuming that the system--bath coupling operator remains the projected flux operator \(\sigma_z\). The key point is that the proof of Theorem~1 uses the explicit form of the noise spectrum only through the existence of a positive global lower bound for the total symmetrized spectrum. The DSS part of the proof, namely the lower bound on \(\sum_k|g_+^{[k]}|^2\), is independent of the detailed spectral model. Therefore, if a modified noise model still satisfies \(\tilde S_{\rm tot}(\omega)\ge S_{\min}^{\rm mod}>0,\) where \(S_{\min}^{\rm mod}\) denotes the global lower-bound constant of the modified symmetrized spectrum,
then the same argument gives \(\gamma_1\ge S_{\min}^{\rm mod} \frac{\omega_d^2}{\omega_d^2+\Delta^2},\)
and hence \(T_1\le \frac{\omega_d^2+\Delta^2}{\omega_d^2 S_{\min}^{\rm mod}}.\)
Thus, changing the spectrum replaces the spectral lower-bound constant in Theorem~1 by the corresponding lower-bound constant of the modified spectrum.

A useful sufficient condition is that the total symmetrized spectrum is lower bounded by two positive power-law contributions with opposite frequency trends. 
Suppose that the total symmetrized spectrum obeys
\begin{equation}
    \tilde S_{\rm tot}(\omega)\ge\frac{a}{|\omega|^{n_1}}+b|\omega|^{n_2},
    \qquad
    a,b>0,\quad n_1,n_2>0
    \label{lower_total}
\end{equation}
Because the first term is large as \(|\omega|\to0\) and the second term is large as \(|\omega|\to\infty\), the right-hand side has a positive global minimum.
The stationary point of $S_{\rm tot}(\omega)$ is obtained by differentiating it with respect to \(|\omega|\):
\(-n_1a|\omega|^{-n_1-1}+n_2b|\omega|^{n_2-1}=0.\)
Equivalently,
\(|\omega|^{n_1+n_2}=\frac{n_1a}{n_2b}.\)
At this point, \(n_1\frac{a}{|\omega|^{n_1}}=n_2b|\omega|^{n_2}\), so substituting the stationary condition back into the right-hand side gives the global lower bound
\begin{equation}
    \tilde S_{\rm tot}(\omega)
    \ge
(n_1+n_2)n_1^{-\frac{n_1}{n_1+n_2}}
n_2^{-\frac{n_2}{n_1+n_2}}
a^{\frac{n_2}{n_1+n_2}}
b^{\frac{n_1}{n_1+n_2}}.
\end{equation}

\emph{Case 1: White-noise floor.}
A simple limiting case is a white-noise floor. If \(\tilde S_{\rm tot}(\omega)=S_0+\tilde S_{\rm col}(\omega),
\;
S_0>0,\,
\tilde S_{\rm col}(\omega)\ge0,\)
where \(S_0\) is the strength of the white-noise floor and \(\tilde S_{\rm col}(\omega)\) is a non-negative colored contribution, then \(\tilde S_{\rm tot}(\omega)\ge S_0\), so one can directly take \(S_{\min}^{\rm mod}=S_0\).

\emph{Case 2: \(1/f\) flux noise plus Ohmic noise.}
For an ohmic contribution coupled through the same \(\sigma_z\) operator, we write \(S_{\rm ohm}(\omega)=A_o|\omega|\kappa(\omega,\mathcal T),\; A_o>0.\)
Here \(A_o\) sets the strength of the ohmic noise component. The corresponding symmetrized spectrum is
$\tilde S_{\rm ohm}(\omega)
=
S_{\rm ohm}(\omega)+S_{\rm ohm}(-\omega)=
A_o|\omega|
\left[
\kappa(\omega,\mathcal T)+\kappa(-\omega,\mathcal T)
\right]
=
A_o|\omega|
\left|
\coth\left(\frac{\hbar\omega}{2k_B\mathcal T}\right)
\right|
\ge
A_o|\omega|$.
Therefore, if the \(1/f\) flux-noise contribution is combined with this ohmic component, the symmetrized spectrum satisfies
\(
\tilde S_{1/f+{\rm ohm}}(\omega)
\ge
\frac{4\pi A_f^2}{|\omega|}+A_o|\omega|.\)
This is a special case of Eq.~\eqref{lower_total}, with \(n_1=n_2=1\), \(a=4\pi A_f^2\), and \(b=A_o\). Therefore,
\(\tilde S_{1/f+{\rm ohm}}(\omega)\ge 2(4\pi A_f^2)^{1/2}A_o^{1/2}=4\sqrt{\pi}\,A_f\sqrt{A_o}.\)

\emph{Case 3: \(1/f\) flux noise plus dielectric loss and Ohmic noise.}
In this case, the total symmetrized spectrum satisfies
\(\tilde S_{\rm tot}(\omega)
\ge
\frac{4\pi A_f^2}{|\omega|}+A_d\left(\frac{\hbar\omega}{2\pi}\right)^2+A_o|\omega|.\)
Since \(A_o|\omega|\ge0\), the lower bound for the \(1/f\) flux-noise plus dielectric-loss spectrum derived in the proof of Theorem~1 remains valid:
\(\tilde S_{\rm tot}(\omega)
\ge 3(A_f^4A_d\hbar^2)^{1/3}.\)
On the other hand, since the dielectric-loss term is also non-negative, the \(1/f\) flux-noise plus ohmic lower bound derived above also applies:
\(\tilde S_{\rm tot}(\omega)
\ge 4\sqrt{\pi}\,A_f\sqrt{A_o}.\)
Therefore, the total symmetrized spectrum has the valid global lower bound
\(\tilde S_{\rm tot}(\omega)
\ge
\max\left[
3(A_f^4A_d\hbar^2)^{1/3},\,
4\sqrt{\pi}\,A_f\sqrt{A_o}
\right].\)
This lower bound is not necessarily the exact global minimum of the three-term spectrum, but it is sufficient to preserve the positive spectral lower-bound condition required in the proof of Theorem~1. 

\emph{Case 4: Generalized \(1/f^\eta\) flux noise plus dielectric loss.}
The same reasoning also applies to generalized non-\(1/f\) flux noise. In this case, the spectrum can be written as
\(S_\eta(\omega)=A_\eta^2\left|\frac{2\pi}{\omega}\right|^\eta + \kappa(\omega,\mathcal T)A_d
\left(\frac{\hbar\omega}{2\pi}\right)^2,
\; \eta>0.\)
Here \(A_\eta\) and \(\eta\) characterize the amplitude and exponent of the generalized \(1/f^\eta\) flux-noise spectrum. Using \(\kappa(\omega,\mathcal T)+\kappa(-\omega,\mathcal T)\ge1\), the symmetrized spectrum satisfies
\(\tilde S_\eta(\omega)
\ge 2A_\eta^2\left|\frac{2\pi}{\omega}\right|^\eta + A_d\left(\frac{\hbar\omega}{2\pi}\right)^2. \)
This is of the special form
\(\frac{a}{|\omega|^{n_1}}+b|\omega|^{n_2},\)
with
\(n_1=\eta,\;
n_2=2,\;
a=2A_\eta^2(2\pi)^\eta,\;
b=\frac{A_d\hbar^2}{4\pi^2}.\)
Therefore, the positive lower-bound of \(\tilde{S}_{\eta}(\omega)\) is
\(S_{\min}^{(\eta)}
=(\eta+2)\eta^{-\frac{\eta}{\eta+2}}
2^{-\frac{2}{\eta+2}}
\left[2A_\eta^2(2\pi)^\eta\right]^{\frac{2}{\eta+2}}
\left(\frac{A_d\hbar^2}{4\pi^2}\right)^{\frac{\eta}{\eta+2}}.\)
}

{
\section{Double-DSS Construction and Dephasing under AC-Amplitude Noise}
\label{double_DSS}

\subsection{Numerical construction of double-DSS points}
\label{subsec:supp_double_dss_construction}

In the main text, the construction of double-DSS points is formulated as a one-dimensional constrained root search along the DSS curve. Here we describe the numerical procedure. For a selected working point on the PF, we fix its normalized nonzero Fourier coefficients \(\{p_k\}_{k\neq0}\) and the drive frequency \(\omega_d\), and vary only the two physical control parameters \(\tilde{\varphi}_{dc}\) and \(\tilde{\varphi}_{ac}\). In this local two-dimensional plane, the Floquet quasienergy gap is written as \(\Omega=\Omega(\tilde{\varphi}_{dc},\tilde{\varphi}_{ac})\), and we define $F_{dc}(\tilde{\varphi}_{dc},\tilde{\varphi}_{ac})=\frac{\partial\Omega}{\partial\tilde{\varphi}_{dc}},
\;
F_{ac}(\tilde{\varphi}_{dc},\tilde{\varphi}_{ac})=\frac{\partial\Omega}{\partial\tilde{\varphi}_{ac}}.$
The double-DSS conditions are $F_{dc}(\tilde{\varphi}_{dc},\tilde{\varphi}_{ac})=0,\; F_{ac}(\tilde{\varphi}_{dc},\tilde{\varphi}_{ac})=0.$

The numerical search is performed within the physical window $\tilde{\Phi}_{dc}/\Phi_0\in[0.521,0.523],\; \tilde{\Phi}_{ac}/\Phi_0\in[0.070,0.080]$.
We first discretize the \(\tilde{\varphi}_{ac}\) direction as $\tilde{\varphi}_{ac}^{(1)}, \tilde{\varphi}_{ac}^{(2)},
\ldots, \tilde{\varphi}_{ac}^{(M)}.$
For each fixed \(\tilde{\varphi}_{ac}^{(m)}\), we solve $F_{dc}\!\left(\tilde{\varphi}_{dc},\tilde{\varphi}_{ac}^{(m)}\right)=0$ along the \(\tilde{\varphi}_{dc}\) direction.
Specifically, we sample \(F_{dc}\) over the \(\tilde{\varphi}_{dc}\) interval and search for a sign-changing bracket. If two neighboring samples \(\tilde{\varphi}_{dc}^{(i)}\) and \(\tilde{\varphi}_{dc}^{(i+1)}\) satisfy
$F_{dc}\!\left(\tilde{\varphi}_{dc}^{(i)},\tilde{\varphi}_{ac}^{(m)}\right)
F_{dc}\!\left(\tilde{\varphi}_{dc}^{(i+1)},\tilde{\varphi}_{ac}^{(m)}\right)<0,$
then the intermediate value theorem guarantees at least one root satisfying \(F_{dc}=0\) in this interval.
Solving this one-dimensional root problem gives a numerical root \(\tilde{\varphi}_{dc}^{(m)}\), which satisfies $F_{dc}\!\left(\tilde{\varphi}_{dc}^{(m)},\tilde{\varphi}_{ac}^{(m)}\right)=0.$
Repeating this procedure for all \(\tilde{\varphi}_{ac}^{(m)}\) gives the discrete sampled points $\left\{
\left(\tilde{\varphi}_{dc}^{(m)},\tilde{\varphi}_{ac}^{(m)}\right)
\right\}_{m}$ on the local DSS curve \(F_{dc}=0\).

If, near the local curve \(F_{dc}=0\), the condition $\frac{\partial F_{dc}}{\partial\tilde{\varphi}_{dc}}\neq0$ is satisfied, the implicit function theorem implies that the solution set of \(F_{dc}=0\) is locally a smooth curve that can be parameterized by \(\tilde{\varphi}_{ac}\). Equivalently, there exists a local smooth function \(f\) such that
\begin{equation}
\tilde{\varphi}_{dc}=f(\tilde{\varphi}_{ac}),
\qquad
F_{dc}\!\left(f(\tilde{\varphi}_{ac}),\tilde{\varphi}_{ac}\right)=0.
\end{equation}
We then evaluate \(F_{ac}\) only at the sampled points that already satisfy \(F_{dc}=0\), and define $G(\tilde{\varphi}_{ac}^{(m)})=F_{ac}\!\left(\tilde{\varphi}_{dc}^{(m)},\tilde{\varphi}_{ac}^{(m)}\right).$
Here \(G\) denotes the value of \(F_{ac}\) restricted to the DSS curve. Numerically, \(F_{ac}\) is evaluated by a central finite difference:
\begin{equation}
F_{ac}(\tilde{\varphi}_{dc},\tilde{\varphi}_{ac})
\approx
\frac{
\Omega(\tilde{\varphi}_{dc},\tilde{\varphi}_{ac}+h)
-
\Omega(\tilde{\varphi}_{dc},\tilde{\varphi}_{ac}-h)
}{2h}.
\end{equation}
If two neighboring DSS samples satisfy $G(\tilde{\varphi}_{ac}^{(m)})G(\tilde{\varphi}_{ac}^{(m+1)})<0,$
then, by the intermediate value theorem, there exists a root \(\tilde{\varphi}_{ac}^{*}\) in the interval \((\tilde{\varphi}_{ac}^{(m)},\tilde{\varphi}_{ac}^{(m+1)})\) such that $G(\tilde{\varphi}_{ac}^{*})=0.$
To locate this root, each new value of \(\tilde{\varphi}_{ac}\) within the sign-changing interval is first mapped to the DSS curve by solving \(F_{dc}(\tilde{\varphi}_{dc},\tilde{\varphi}_{ac})=0\). The resulting \(\tilde{\varphi}_{dc}=f(\tilde{\varphi}_{ac})\) is then used to evaluate \(G(\tilde{\varphi}_{ac})=F_{ac}\!\left(f(\tilde{\varphi}_{ac}),\tilde{\varphi}_{ac}\right)\).
Thus the second-level search is always carried out on the same local DSS curve and remains constrained by \(F_{dc}=0\). The final point $\left(\tilde{\varphi}_{dc}^{*},\tilde{\varphi}_{ac}^{*}\right)=\left(f(\tilde{\varphi}_{ac}^{*}),\tilde{\varphi}_{ac}^{*}\right)$
satisfies
\begin{equation}
F_{dc}\!\left(\tilde{\varphi}_{dc}^{*},\tilde{\varphi}_{ac}^{*}\right)=0,
\qquad
F_{ac}\!\left(\tilde{\varphi}_{dc}^{*},\tilde{\varphi}_{ac}^{*}\right)=0.
\end{equation}
This point is therefore a double-DSS.

If, for some fixed \(\tilde{\varphi}_{ac}\), no sign-changing bracket of \(F_{dc}\) is found along the \(\tilde{\varphi}_{dc}\) direction within the prescribed local window, the corresponding DSS sampled point cannot be obtained. If the value of \(G\) along the local DSS curve \(\tilde{\varphi}_{dc}=f(\tilde{\varphi}_{ac})\) does not change sign throughout the \(\tilde{\varphi}_{ac}\) window, then this local window does not contain a double-DSS constructed by the present method. If multiple roots of \(F_{dc}=0\) exist for some fixed \(\tilde{\varphi}_{ac}\), we only retain roots that lie on the same locally connected DSS curve as the selected PF point; if such a local connection cannot be determined, that PF point does not yield a valid double-DSS.

\subsection{AC-amplitude-noise dephasing model for double-DSS validation}
\label{subsec:supp_ac_noise_dephasing}

This subsection derives the pure-dephasing contribution from AC amplitude noise. According to Eq.~(3) in the main text, the projected two-level Hamiltonian is
\begin{equation}
H_q(t)=\frac{\Delta}{2}\sigma_x+\frac{B}{2}\sigma_z+A P(t)\sigma_z,
\qquad
A=E_L\varphi_{ac}\tilde{\phi}_{ge}.
\end{equation}
The part depending on the AC amplitude \(\varphi_{ac}\) is \(E_L\varphi_{ac}\tilde{\phi}_{ge}P(t)\sigma_z\). Therefore, $\frac{\partial H_q(t)}{\partial\varphi_{ac}} = E_L\tilde{\phi}_{ge}P(t)\sigma_z .$
When the AC amplitude fluctuates as \(\varphi_{ac}\rightarrow\varphi_{ac}+\delta\varphi_{ac}(t)\), the first-order perturbation to the Hamiltonian is
\begin{equation}
\delta H(t)
=
\delta\varphi_{ac}(t)
\frac{\partial H_q(t)}{\partial\varphi_{ac}}
=
E_L\tilde{\phi}_{ge}\,\delta\varphi_{ac}(t)P(t)\sigma_z.
\end{equation}
The overall factor \(E_L\tilde{\phi}_{ge}\) can be absorbed into the noise spectrum. Hence, in what follows, we only keep the clean system operator \(P(t)\sigma_z\).

We now project this AC-amplitude-noise channel onto the Floquet basis $|\omega_\pm(t)\rangle$. As in the derivation of \(g_z^{[k]}\) in Eq.~(10) of the main text, the longitudinal pure-dephasing channel is determined by the difference between the diagonal matrix elements of the noise-coupling operator in the two Floquet modes. Replacing \(\sigma_z\) in the main-text expression by the AC-amplitude-noise coupling operator \(P(t)\sigma_z\), we obtain
\begin{equation}
g_{z,ac}^{[m]}
=
\frac{1}{2T}
\int_0^T dt\,e^{im\omega_d t}
\left[
\left\langle \omega_+(t)\left|
P(t)\sigma_z
\right|\omega_+(t)\right\rangle
-
\left\langle \omega_-(t)\left|
P(t)\sigma_z
\right|\omega_-(t)\right\rangle
\right],
\label{eq:gzac_def}
\end{equation}
where the normalization convention is the same as that used for \(g_z^{[k]}\) in the main text. Using the coefficients \(g_z^{[k]}\) introduced in Eq.~(10), \(g_{z,ac}^{[m]}\) can be written conveniently in terms of the original dephasing harmonics. From the Fourier expansion of \(g_z^{[k]}\), we have $\left\langle \omega_+(t)\left|\sigma_z\right|\omega_+(t)\right\rangle
-
\left\langle \omega_-(t)\left|\sigma_z\right|\omega_-(t)\right\rangle
=
\sum_{\ell\in\mathbb Z}g_z^{[\ell]}e^{-i\ell\omega_d t}.$
Together with \(P(t)=\sum_n p_n e^{in\omega_d t}\), this gives
\begin{equation}
\left[
\left\langle \omega_+(t)\left|
P(t)\sigma_z
\right|\omega_+(t)\right\rangle
-
\left\langle \omega_-(t)\left|
P(t)\sigma_z
\right|\omega_-(t)\right\rangle
\right]
=
\sum_{n,\ell}p_n g_z^{[\ell]}e^{-i(\ell-n)\omega_d t}.
\end{equation}
Comparing this expression with the Fourier component in Eq.~\eqref{eq:gzac_def}, we obtain $g_{z,ac}^{[m]} = \sum_n p_n g_z^{[m+n]}.$
In particular, the zero-frequency component is $g_{z,ac}^{[0]}=\sum_n p_n g_z^{[n]}.$

The leading contribution from low-frequency \(1/f\) AC amplitude noise is governed by the zero-frequency component \(g_{z,ac}^{[0]}\). We next show that \(g_{z,ac}^{[0]}\) is equal to the first derivative of the quasienergy gap with respect to the AC amplitude. Differentiating the Floquet eigenvalue equation $\left[H_q(t)-i\frac{\partial}{\partial t}\right]|\omega_\pm(t)\rangle=\epsilon_\pm|\omega_\pm(t)\rangle$
with respect to \(\varphi_{ac}\), we obtain
\begin{equation}
\frac{\partial H_q(t)}{\partial\varphi_{ac}}|\omega_\pm(t)\rangle
+
\left[H_q(t)-i\frac{\partial}{\partial t}\right]
\frac{\partial|\omega_\pm(t)\rangle}{\partial\varphi_{ac}}
=
\frac{\partial\epsilon_\pm}{\partial\varphi_{ac}}
|\omega_\pm(t)\rangle
+
\epsilon_\pm
\frac{\partial|\omega_\pm(t)\rangle}{\partial\varphi_{ac}} .
\end{equation}
Multiplying from the left by \(\langle\omega_\pm(t)|\) and averaging over one period gives
\begin{equation}
\frac{\partial\epsilon_\pm}{\partial\varphi_{ac}}
=
\frac{1}{T}
\int_0^T dt\,
\left\langle \omega_\pm(t)\left|
\frac{\partial H_q(t)}{\partial\varphi_{ac}}
\right|\omega_\pm(t)\right\rangle .
\end{equation}
Therefore, for the quasienergy gap \(\Omega=\epsilon_+-\epsilon_-\), we have
\begin{equation}
\frac{\partial\Omega}{\partial\varphi_{ac}}
=
\frac{1}{T}\int_0^T dt\,\left[\left\langle \omega_+(t)\left|\frac{\partial H_q(t)}{\partial\varphi_{ac}}\right|\omega_+(t)\right\rangle
-
\left\langle \omega_-(t)\left|\frac{\partial H_q(t)}{\partial\varphi_{ac}}\right|\omega_-(t)\right\rangle\right].
\end{equation}
Substituting \(\partial H_q(t)/\partial\varphi_{ac}\) as \(P(t)\sigma_z\), yields $g_{z,ac}^{[0]} = \frac{\partial\Omega}{\partial{\varphi}_{ac}}.$
The derivation above is carried out using the original waveform \(P(t)\) and amplitude \(\varphi_{ac}\). In the normalized waveform representation used in the main text, the external flux can equivalently be written as \(\phi_{\mathrm{ext}}(t)=\tilde{\phi}_{dc}+\tilde{\phi}_{ac}\tilde P(t)\), with \(\max_t|\tilde P(t)|=1\). The two-level Hamiltonian has the same structure in this representation, with \(\varphi_{ac}\) and \(P(t)\) replaced by \(\tilde{\varphi}_{ac}\) and \(\tilde P(t)\), respectively. Hence, the following expressions also apply in the normalized representation. We use \(\tilde{\varphi}_{ac}\) and \(\tilde P(t)\) below, consistent with the notation in the main text.

For low-frequency \(1/f\) AC amplitude noise, we take $S_{ac}(\omega)=A_{f,ac}^{2}\left|\frac{2\pi}{\omega}\right|.$
As in the treatment of the zero-frequency \(1/f\) flux-noise term in Eq.~(13) of the main text, the dominant low-frequency dephasing contribution is determined by the zero-frequency longitudinal coupling coefficient of the corresponding noise channel. For AC amplitude noise, this gives the low-frequency contribution $\left|
\frac{\partial\Omega}{\partial\tilde{\varphi}_{ac}}
\right|
A_{f,ac}
\sqrt{2|\ln(\omega_{ir}t_m)|}.$ 
Since \(V_{ac}(t)=\tilde P(t)\sigma_z\) is itself periodically modulated, AC amplitude noise also enters the finite-frequency pure-dephasing channels through its Floquet sidebands. We treat these sideband contributions in the same way as the finite-frequency Floquet-harmonic terms derived from Eq.~(10) to Eq.~(13) in the main text. Replacing the dc-noise harmonic coefficients \(g_z^{[k]}\) by the AC-noise harmonic coefficients \(g_{z,ac}^{[k]}\), the finite-frequency contribution from the AC-amplitude-noise channel is $\sum_{k\neq0}
|g_{z,ac}^{[k]}|^2
S_{ac}(k\omega_d).$
The pure-dephasing rate associated with the AC amplitude-noise channel is then
\begin{equation}
\gamma_{\phi}^{ac}
=
\left|
\frac{\partial\Omega}{\partial\tilde{\varphi}_{ac}}
\right|
A_{f,ac}
\sqrt{2|\ln(\omega_{ir}t_m)|}
+
\sum_{k\neq0}
|g_{z,ac}^{[k]}|^2
S_{ac}(k\omega_d).
\end{equation}
Here \(A_{f,ac}\) denotes the low-frequency \(1/f\) noise amplitude of AC amplitude fluctuations. In the numerical comparison, we set \(A_{f,ac}=A_f\) as a baseline assumption, so that the DC flux-noise and AC amplitude-noise channels are compared on the same low-frequency noise scale.

Finally, we denote the DC flux-noise pure-dephasing rate in Eq.~(13) of the main text by \(\gamma_{\phi}^{dc}\). Including the AC-amplitude-noise channel derived above, the effective pure-dephasing rate used for the comparison is $\gamma_{\phi}^{dc+ac}=\gamma_{\phi}^{dc}+\gamma_{\phi}^{ac},$
and the corresponding effective pure-dephasing time is $T_{\phi,dc+ac}={1}/{\gamma_{\phi}^{dc+ac}}.$
This linear addition overestimates the total pure-dephasing rate and therefore underestimates the effective pure-dephasing time, making it a conservative definition.
}

{
\section{Floquet leakage-channel analysis and gate validation}
\label{supsec:gate_supplement}

\subsection{Frequency-domain form of the leakage-channel penalty}
\label{supsec:channel_regularizer}

\begin{figure*}[t]
        \centering
    \includegraphics[width=\linewidth]{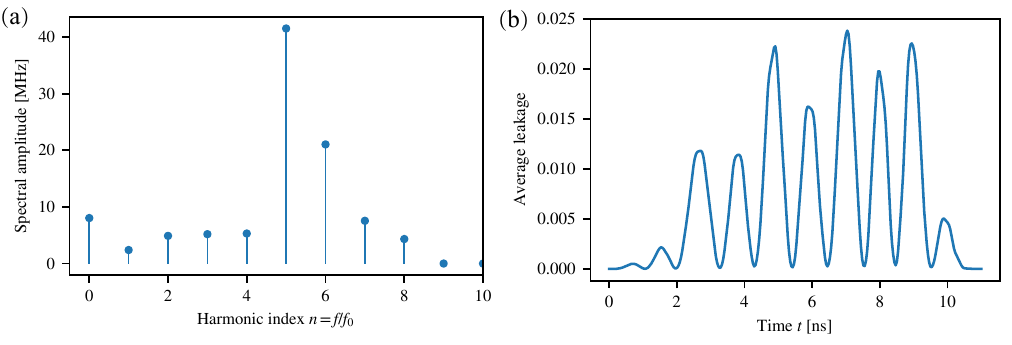}
    \caption{
Additional numerical results for the single-qubit \(X\) gate.
(a) Spectrum of the optimized Floquet-logical control pulse \(\Omega_F(t)\). The horizontal axis is the harmonic index \(n=f/f_0\), where \(f_0=1/T_g\) is the fundamental frequency set by the total gate duration \(T_g\). The vertical axis gives the corresponding spectral amplitude in MHz.
(b) Time-dependent average leakage \(L_{\rm avg}(t)\) during the full multilevel gate evolution.
}
    \label{fig:supp_x_gate_results}
\end{figure*}\label{fig:supp_iswap_gate_results}
\begin{figure*}[t]
    \centering
    \includegraphics[width=\linewidth]{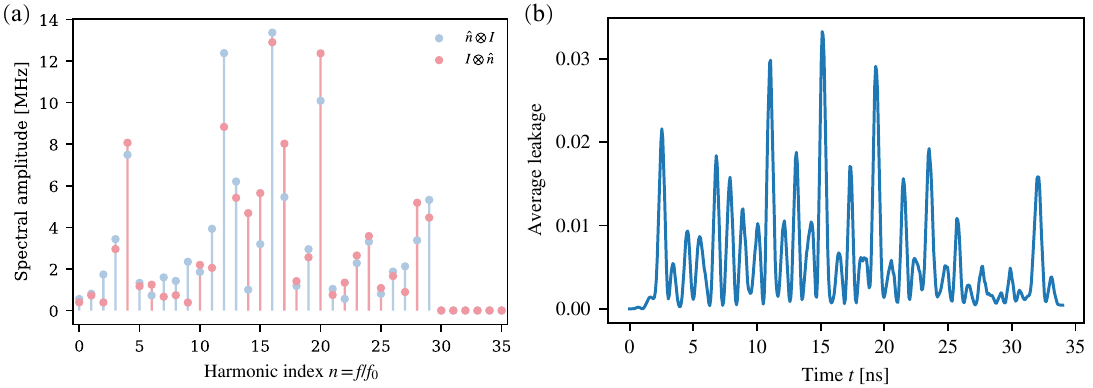}
    \caption{
Additional numerical results for the two-qubit \(\sqrt{i{\rm SWAP}}\) gate.
(a) Spectra of the two optimized local Floquet-logical control pulses. The two spectra correspond to the local charge-control channels \(\hat n\otimes I\) and \(I\otimes\hat n\), respectively. 
(b) Time-dependent average leakage \(L_{\rm avg}(t)\) during the full multilevel two-qubit evolution.
}
    \label{fig:supp_iswap_gate_results}
\end{figure*}

This subsection gives a frequency-domain interpretation of the leakage-channel penalty used in the gate optimization. Starting from the first-order Dyson amplitude matrix
\(A_{\rm ch}^{(1)}=-i\sum_\mu\int_0^{T_g}f_\mu(t)P_{\rm ch}\widetilde n_\mu(t)F\,dt\),
we show that \(\mathcal L_{\rm ch}\) suppresses the spectral weight of the control pulses at the selected Floquet-dressed leakage frequencies. We denote the set of Floquet computational-state labels by \(C\), with \(C=\{+,-\}\) for a single-qubit gate and \(C=\{++,+-,-+,--\}\) for a two-qubit gate. We denote the selected high-level non-computational states included in the leakage penalty by \(C^\perp\), and write
\(P_{\rm ch}=\sum_{\ell\in C^\perp}|\omega_\ell(0)\rangle\langle\omega_\ell(0)|.\) 
For \(a\in C\) and \(\ell\in C^\perp\), the first-order leakage amplitude from the Floquet computational state \(|\omega_a(0)\rangle\) to the high-level state \(|\omega_\ell(0)\rangle\) is defined as
\begin{equation}
a_{\ell a}^{(1)}(T_g)=-i\sum_\mu\int_0^{T_g} f_\mu(t)\langle\omega_\ell(0)|\widetilde n_\mu(t)|\omega_a(0)\rangle dt .
\label{eq:supp_first_order_channel_amp}
\end{equation}
This amplitude is the matrix element of \(A_{\rm ch}^{(1)}\) associated with leakage from the computational state \(a\) to the high-level state \(\ell\). Since the selected high-level channels are weighted equally in this work, the leakage penalty can be written as
\begin{equation}
\mathcal L_{\rm ch}
=\frac{1}{d}\left\|A_{\rm ch}^{(1)}\right\|_F^2
=\frac{1}{d}\sum_{a\in C}\sum_{\ell\in C^\perp}\left|a_{\ell a}^{(1)}(T_g)\right|^2 .
\label{eq:supp_Lch_channel_sum}
\end{equation}

We now rewrite \(a_{\ell a}^{(1)}(T_g)\) in the frequency domain. Using \(\widetilde n_\mu(t)=U_q^\dagger(t)\hat n_\mu U_q(t),\) together with the Floquet relation \(U_q(t)|\omega_a(0)\rangle=e^{-i\epsilon_a t}|\omega_a(t)\rangle,\)
we obtain $\langle\omega_\ell(0)| \widetilde n_\mu(t)|\omega_a(0)\rangle=e^{i(\epsilon_\ell-\epsilon_a)t}\langle\omega_\ell(t)|\hat n_\mu|\omega_a(t)\rangle .$ Since the Floquet modes are periodic, the matrix element \(\langle\omega_\ell(t)|\hat n_\mu|\omega_a(t)\rangle\) is also periodic and can be expanded as $\langle\omega_\ell(t)|\hat n_\mu|\omega_a(t)\rangle=\sum_mn_{\mu,\ell a}^{(m)}e^{-im\omega_d t}.$ Therefore, Eq.~\eqref{eq:supp_first_order_channel_amp} becomes
\begin{equation}
\begin{split}
    a_{\ell a}^{(1)}(T_g)
    =-i\sum_\mu\sum_m n_{\mu,\ell a}^{(m)}\int_0^{T_g} f_\mu(t)e^{i(\epsilon_\ell-\epsilon_a-m\omega_d)t}dt 
    \equiv -i\sum_\mu\sum_m
n_{\mu,\ell a}^{(m)}
\widehat f_{\mu,T_g}
(\epsilon_\ell-\epsilon_a-m\omega_d).
\end{split}
\label{eq:supp_leakage_amp_frequency}
\end{equation}
Here $\widehat f_{\mu,T_g}(\Omega)=\int_0^{T_g}f_\mu(t)e^{i\Omega t}dt$ is the finite-time Fourier transform of the control pulse over the gate duration.
Substituting Eq.~\eqref{eq:supp_leakage_amp_frequency} into Eq.~\eqref{eq:supp_Lch_channel_sum} gives
\begin{equation}
\mathcal L_{\rm ch}=\frac{1}{d}\sum_{a\in C}\sum_{\ell\in C^\perp}\left|\sum_\mu\sum_m n_{\mu,\ell a}^{(m)}\widehat f_{\mu,T_g}(\epsilon_\ell-\epsilon_a-m\omega_d)\right|^2 .
\label{eq:supp_Lch_frequency}
\end{equation}
This is the frequency-domain form of \(\mathcal L_{\rm ch}\). It shows that the leakage penalty depends not only on the overall bandwidth of the pulse, but also on the Floquet quasienergy differences, drive sidebands, and the Fourier components of the control-operator matrix elements between Floquet modes.

In the numerical optimization, the control pulse is parameterized using a finite number of Fourier components,
\(f_\mu(t)=\sum_r\left[a_{\mu r}\cos(\nu_r t)+b_{\mu r}\sin(\nu_r t)\right],\)
where \( \nu_r=\frac{2\pi r}{T_g}\), and \(a_{\mu r}\) and \(b_{\mu r}\) are trainable parameters. With this parameterization, \(\widehat f_{\mu,T_g}(\Omega)\) can be written as a sum of the window responses of the individual Fourier components at frequency \(\Omega\):
\begin{equation}
\widehat f_{\mu,T_g}(\Omega)
=\sum_r a_{\mu r}\frac{\mathcal I_{T_g}(\Omega+\nu_r)+\mathcal I_{T_g}(\Omega-\nu_r) }{2}
+\sum_r b_{\mu r}\frac{\mathcal I_{T_g}(\Omega+\nu_r)-\mathcal I_{T_g}(\Omega-\nu_r)}{2i},
\label{eq:supp_finite_time_ft_fourier}
\end{equation}
where $\mathcal I_{T_g}(x)=\int_0^{T_g} e^{ixt}dt=e^{ixT_g/2}T_g\,{\rm sinc}\left(\frac{xT_g}{2}\right).$
For \(x\approx0\), one has \(|\mathcal I_{T_g}(x)|\approx T_g\), whereas for \(|x|T_g\gg1\) the response is suppressed by the sinc function. Thus, if a Floquet-dressed leakage frequency \(\Omega_{\ell a}^{(m)} = \epsilon_\ell-\epsilon_a-m\omega_d\) is close to one of the pulse Fourier frequencies, i.e., \(\Omega_{\ell a}^{(m)}\pm\nu_r\approx0\), the corresponding \(\mathcal I_{T_g}\) term in Eq.~\eqref{eq:supp_finite_time_ft_fourier} can coherently accumulate and enhance the first-order leakage amplitude. Conversely, if the optimized pulse suppresses \(\widehat f_{\mu,T_g}\!\left(\Omega_{\ell a}^{(m)}\right) \)
at the dominant leakage frequencies, the corresponding leakage amplitudes in Eq.~\eqref{eq:supp_leakage_amp_frequency} are reduced. Therefore, \(\mathcal L_{\rm ch}\) can be understood as a frequency-domain penalty targeted at selected Floquet high-level leakage channels.

\begin{figure*}[t]
    \centering
    \begin{tikzpicture}
    \begin{groupplot}[
        group style={
            group size=2 by 1,
            horizontal sep=18mm
        },
        width=0.40\textwidth,
        height=0.28\textwidth,
        scale only axis,
        xmin=4.8,
        xmax=12.2,
        xtick={5,6,7,8,9,10,11,12},
        ymode=log,
        grid=both,
        major grid style={line width=0.35pt, draw=gray!35},
        minor grid style={line width=0.20pt, draw=gray!20},
        tick align=outside,
        tick pos=left,
        axis line style={black!70},
        tick style={black!70},
        label style={font=\small},
        tick label style={font=\small},
        legend cell align=left,
        every axis plot/.append style={line width=1.0pt},
        enlarge x limits=0.03,
    ]

    \nextgroupplot[
        ylabel={$1-F_{\rm avg}$},
        ymin=7e-6,
        ymax=3.5e-3,
    ]
    \node[anchor=north west, font=\small\bfseries]
        at (axis description cs:0.05,0.95) {(a)};

    \addplot[
        color=journalblue,
        mark=o,
        mark size=2.6pt,
        mark options={draw=journalblue, fill=white, line width=0.9pt},
    ] coordinates {
        (5,1.11910566e-05)
        (6,8.78247603e-05)
        (7,9.61200250e-05)
        (8,9.70016921e-05)
        (9,9.831398998e-05)
        (10,9.874328955e-05)
        (11,9.927170698e-05)
        (12,9.993707264e-05)
    };

    \addplot[
        color=journalbrick,
        mark=square*,
        mark size=2.9pt,
        mark options={draw=journalbrick, fill=journalbrick},
    ] coordinates {
        (5,1.185063371000e-04)
        (6,1.390583706000e-04)
        (7,4.201088472000e-04)
        (8,4.588034421000e-04)
        (9,5.739403740000e-04)
        (10,5.718647721000e-04)
        (11,6.188766300000e-04)
        (12,8.485113412000e-04)
    };

    \nextgroupplot[
        ylabel={$L_{\rm avg}$},
        ymin=5e-6,
        ymax=3.5e-3,
        legend style={
            at={(0.98,0.99)},
            anchor=north east,
            draw=none,
            fill=white,
            fill opacity=0.82,
            text opacity=1,
            font=\scriptsize,
            inner xsep=2pt,
            inner ysep=2pt,
            row sep=1pt,
        },
    ]
    \node[anchor=north west, font=\small\bfseries]
        at (axis description cs:0.05,0.95) {(b)};

    \addplot[
        color=journalblue,
        mark=o,
        mark size=2.6pt,
        mark options={draw=journalblue, fill=white, line width=0.9pt},
    ] coordinates {
        (5,7.899767521469e-06)
        (6,9.334184872811e-06)
        (7,9.339522900609e-06)
        (8,9.353225336017e-06)
        (9,9.352414507724e-06)
        (10,9.352713182142e-06)
        (11,9.352593939860e-06)
        (12,9.353127532696e-06)
    };
    \addlegendentry{Single-qubit \(X\)}
    
    \addplot[
        color=journalbrick,
        mark=square*,
        mark size=2.9pt,
        mark options={draw=journalbrick, fill=journalbrick},
    ] coordinates {
        (5,1.849820246902e-05)
        (6,3.775908514081e-05)
        (7,2.968767285747e-04)
        (8,3.305100453601e-04)
        (9,4.446294282618e-04)
        (10,4.421060845681e-04)
        (11,4.894847076393e-04)
        (12,7.182110105011e-04)
    };
    \addlegendentry{Two-qubit \(\sqrt{i{\rm SWAP}}\)}

    \end{groupplot}
    \end{tikzpicture}
    \caption{
    Numerical results at different validation truncation dimensions.
    (a) Gate error \(1-F_{\rm avg}\).
    (b) Average leakage \(L_{\rm avg}\).
    Blue open circles denote the single-qubit \(X\) gate, and brick-red filled squares denote the two-qubit \(\sqrt{i{\rm SWAP}}\) gate.
    }
    \label{fig:supp_truncation_validation}
\end{figure*}

\subsection{Additional numerical results for single- and two-qubit gates}
\label{supsec:numerical_settings}

This subsection collects additional numerical results not shown in the main text, including pulse spectra, time-dependent average leakage \(L_{\rm avg}(t)\), and the dependence of the gate error \(1-F_{\rm avg}\) and average leakage \(L_{\rm avg}\) on the validation truncation dimension.

For the single-qubit \(X\) gate, the additional results are shown in Fig.~\ref{fig:supp_x_gate_results}. Figure~\ref{fig:supp_x_gate_results}(a) shows the spectrum of the optimized pulse \(\Omega_F(t)\), while Fig.~\ref{fig:supp_x_gate_results}(b) shows the time-dependent average leakage \(L_{\rm avg}(t)\).

For the two-qubit \(\sqrt{i{\rm SWAP}}\) gate, the additional results are shown in Fig.~\ref{fig:supp_iswap_gate_results}. Figure~\ref{fig:supp_iswap_gate_results}(a) shows the spectra of the two local control pulses \(\Omega_{F,1}(t)\) and \(\Omega_{F,2}(t)\), and Fig.~\ref{fig:supp_iswap_gate_results}(b) shows the time-dependent average leakage \(L_{\rm avg}(t)\).

To examine the dependence of the optimized pulses on the Hilbert-space truncation, Fig.~\ref{fig:supp_truncation_validation} shows the gate error \(1-F_{\rm avg}\) and the average leakage \(L_{\rm avg}\) evaluated at different validation truncation dimensions. Both vertical axes are plotted on a logarithmic scale. The single-qubit gate remains nearly unchanged as higher levels are included. 
The two-qubit gate is more sensitive to the truncation dimension: both the gate error and the average leakage increase overall as additional higher levels are included, but the validation at \(N=12\) still gives \(F_{\rm avg}\approx 99.92\%\) and \(L_{\rm avg}\approx 7.18\times10^{-4}\).
}


\end{document}